\documentclass[12pt,draftcls,lettersize,journal,onecolumn]{IEEEtran}
\usepackage{amsmath,amssymb,amsfonts}
\usepackage{algorithm}
\usepackage{algcompatible}
\usepackage{array}
\usepackage[caption=false,font=normalsize,labelfont=sf,textfont=sf]{subfig}
\usepackage{textcomp}
\usepackage{stfloats}
\usepackage{url}
\usepackage{verbatim}
\usepackage{graphicx}
\usepackage{cite}
\usepackage{multirow}
\usepackage{tabularx}
\usepackage{xcolor}
\usepackage{soul}
\algnewcommand\INPUT{\item[\textbf{Input:}]}%
\algnewcommand\OUTPUT{\item[\textbf{Output:}]}%

\graphicspath{{./figure/}}

\begin{document}
{\Large \textbf{Notice:} This work has been submitted to the IEEE for possible publication.
Copyright may be transferred without notice, after which this version may no longer be accessible.\par}
\newpage

\title{Deep Learning-aided Parametric Sparse Channel Estimation for Terahertz Massive MIMO Systems}

\author{Jinhong Kim,~\IEEEmembership{Member,~IEEE,}
        Yongjun Ahn,~\IEEEmembership{Member,~IEEE,}
        Seungnyun Kim,~\IEEEmembership{Member,~IEEE,}
        and~Byonghyo Shim,~\IEEEmembership{Senior Member,~IEEE,}%
\thanks{A preliminary version of this paper was presented at the International Conference on Communications (ICC), Seoul, Korea, 16-20 May, 2022~\cite{jhkim2022parametric}.

This work was supported by the Institute of Information and Communications Technology Planning and Evaluation (IITP) Grant funded by the Korea Government (MSIT, Development of Intelligent Wireless Access Technologies) under Grant 2021-0-00972 and the National Research Foundation of Korea (NRF) grant funded by the Korea government (MSIT) (2022M3C1A3099336 and RS-2023-00252789).}

\thanks{Jinhong Kim and Yongjun Ahn are with Samsung Electronics, Seoul 06765, (e-mail: {jinhong3.kim@samsung.com}; {yong\_jun.ahn@samsung.com})

Seungnyun Kim is with the Laboratory for Information and Decision Systems, Massachusetts Institute of Technology, Cambridge, MA 02139 USA (e-mail: {snkim94@mit.edu}).

Byonghyo Shim is with the Department of Electrical and Computer Engineering and the Institute of New Media and Communications, Seoul National University, Seoul 08826, Republic of Korea (e-mail: {bshim@snu.ac.kr}).}
}

\maketitle
\vspace{-4.5em}

\begin{abstract}
\vspace{-1.4em}
Terahertz (THz) communications is considered as one of key solutions to support extremely high data demand in 6G.
One main difficulty of the THz communication is the severe signal attenuation caused by the foliage loss, oxygen/atmospheric absorption, body and hand losses.
To compensate for the severe path loss, multiple-input-multiple-output (MIMO) antenna array-based beamforming has been widely used.
Since the beams should be aligned with the signal propagation path to achieve the maximum beamforming gain, acquisition of accurate channel knowledge, i.e., channel estimation, is of great importance.
An aim of this paper is to propose a new type of deep learning (DL)-based parametric channel estimation technique.
In our work, DL figures out the mapping function between the received pilot signal and the sparse channel parameters characterizing the spherical domain channel.
By exploiting the long short-term memory (LSTM), we can efficiently extract the temporally correlated features of sparse channel parameters and thus make an accurate estimation with relatively small pilot overhead.
From the numerical experiments, we show that the proposed scheme is effective in estimating the near-field THz MIMO channel in THz downlink environments.
\end{abstract}
\vspace{-1.6em}
\begin{IEEEkeywords}
\vspace{-1.5em}
time-varying channel estimation, near-field transmission, terahertz (THz) communication system, long short-term memory (LSTM)
\vspace{-2.4em}
\end{IEEEkeywords}

\newpage
\section{Introduction}

\IEEEPARstart{T}{erahertz} (THz) communication has stimulated great deal of interest as an essential technology to meet the ever-increasing demand for higher data rates in 6G in recent years~\cite{jiang2021road,khan20206g,bariah2020prospective,han2022terahertz}.
Exploiting the plentiful spectrum resources in the THz frequency band ($0.1\,\text{THz}\sim 10\,\text{THz}$), THz communication can support way higher data rate than the conventional sub-6GHz and millimeter-wave wireless communication systems can offer.
However, handling of THz spectrum is challenging since the signal power is severely degraded by the high diffraction and penetration loss, atmospheric absorption, and rain attenuation~\cite{wu2022globecom}.
To overcome the problem, the beamforming technique using the multiple-input-multiple-output (MIMO) antenna arrays has been widely used~\cite{ning2021terahertz,park2017expectation}.
Since the beamforming gain is maximized only when the transmit beams are aligned with the channel propagation paths, obtaining an accurate channel information is crucial for the THz MIMO systems.

Traditionally, linear estimation techniques such as least squares (LS) and linear minimum mean square error (LMMSE) estimators have been popularly used for the channel estimation of the MIMO systems~\cite{liu2014channel, coleri2002channel, choi2017compressed, wan2019compressive}.
Potential problem of the conventional approaches is that the amount of pilot resources needed for the channel estimation are proportional to the number of the transmit antennas.
This problem is even more serious in the THz communication systems since the required number of transmit antennas to achieve the beamforming gain is quite large (hundred to thousand elements).
For example, if the number of transmit and receive antennas are $N_{T}=64$ and $N_{R}=8$, respectively, then the base station (BS) has to transmit at least 4 resource blocks (RBs) ($12\times 14$ resources in each RB of 5G NR) just for the pilot transmission, occupying almost $25\%$ of a subframe in 5G NR under the 240 kHz subcarrier spacing.

In order to reduce the pilot transmission overhead associated with the channel estimation, compressed sensing (CS)-based channel estimation technique has been proposed~\cite{choi2017compressed, wan2019compressive, kwon2014multipath}.
Essence of this approach is to use the property that the number of effective paths is at most a few (line-of-sight (LoS) and $1\sim 2$ non-line-of-sight (NLoS) paths) in the THz channel.
Under this condition, one can readily convert the channel estimation problem into the sparse recovery problem on the angular domain.
Since the premise of the CS technique is that the sparse channel can be recovered with measurements whose size being proportional to $O(k\log (n/k))$ where $k$ is the number of propagation paths and $n$ is dimension of quantized angles, the required pilot resources for the channel estimation can be reduced considerably.

While the CS-based channel estimation is effective to some extent, the benefit will vanish when the channel exhibits the near-field effect characteristics.
Since the wavefront of the THz electromagnetic signal is spherical, conventional angular domain channel model obtained by the planar wavefront model in far-field will not be accurate and, in fact, make a severe mismatch in most practical scenarios~\cite{lee2022globecom}.
This mismatch between the real and modeled channels will clearly lead to the severe degradation in the channel estimation performance so that a new channel model and estimation technique suited for the near-field THz channel is needed.

An aim of this paper is to propose a novel channel estimation technique for the THz communication systems.
Exploiting the property that the near-field THz channel can be expressed as a few parameters in the spherical domain, viz., angle of departures (AoDs), angle of arrivals (AoAs), distances, and path gains, the proposed technique, henceforth referred to as \textit{deep sparse time-varying channel estimation} (D-STiCE), estimates the sparse channel parameters and then reconstructs the channel using the obtained parameters.
To estimate the sparse channel parameters in the continuous domain, D-STiCE employs a deep learning (DL) technique, a data-driven learning approach to approximate the desired function.
In a nutshell, the proposed D-STiCE learns the mapping function between the sequential data (in our case, pilot measurements) and the continuous channel parameters.
As a main engine for the task at hand, we exploit the long short-term memory (LSTM), a model specialized for extracting temporally correlated features from the sequential data.
By extracting the temporal correlation of channel parameters, we make a fast yet accurate channel estimation with relatively small amount of pilot resources.

The main contributions of this paper are as follows.
\begin{itemize}
    \item We propose a two-stage channel estimation scheme for THz massive MIMO systems.
    Our approach includes 1) large-scale parameter estimation to identify the geometric channel parameters and 2) small-scale parameter estimation to identify the random variations of the path gain.
    Since the geometric parameters vary much slower than the path gains, we estimate the parameters in different time scales.
    That is, we estimate the angles and distance occasionally and estimate the path gains frequently.
    In doing so, we can avoid the waste of pilot resources required for the frequent estimation of static or quasi-static channel components.
    \item We design a novel DL-based channel estimation tailored for the THz near-field environment.
    In our approach, we perform the nonlinear decomposition of the near-field THz channel to express the channel using a few geometric parameters.
    We exploit LSTM to extract the temporal correlation among geometric parameters in pilot measurements, which is then used to identify the large-scale near-field channel parameters.

    \item To validate the effectiveness of the proposed D-STiCE, we conduct numerical evaluations in the realistic THz massive MIMO environments.
    We demonstrate that D-STiCE outperforms the conventional LS, LMMSE, CS-based channel estimation, and also CNN-based channel estimation technique by a large margin.
    For example, D-STiCE achieves more than $4\,$dB gain in NMSE over the LS and LMMSE in high SNR regime.
    Even when compared to the CS and CNN-based scheme, D-STiCE achieves more than $3\,\text{dB}$ gain in bit error rate (BER).        
\end{itemize}

The rest of this paper is organized as follows.
In Section II, we discuss the near-field THz MIMO-OFDM system and briefly review conventional channel estimation techniques.
In Section III, we present the D-STiCE technique and provide a detailed description of the model.
In Section IV, we discuss various implementation issues.
In Section V, we present simulation results to verify the performance gain of the proposed technique and conclude the paper in Section VI.

\textit{Notations}: Lower and upper case symbols are used to denote vectors and matrices, respectively.
The superscripts $(\cdot)^{\textrm{T}}$, $(\cdot)^{\textrm{H}}$, and $(\cdot)^{+}$ denote the transpose, hermitian transpose, and pseudo-inverse, respectively.
$\lVert\mathbf{x}\rVert$ is used as the Euclidean norm of a vector $\mathbf{x}$.
$\text{vec}(\mathbf{X})$ denotes the vectorization of $\mathbf{X}$ and $\text{diag}(\mathbf{x})$ denotes a diagonal matrix whose diagonal elements are $\mathbf{x}$.
In addition, $\mathbf{X}_1 \otimes \mathbf{X}_2$ denote the Kronecker products of $\mathbf{X}_1$ and $\mathbf{X}_2$.

\begin{figure*}[t]
	\centering
	\includegraphics[width=\linewidth]{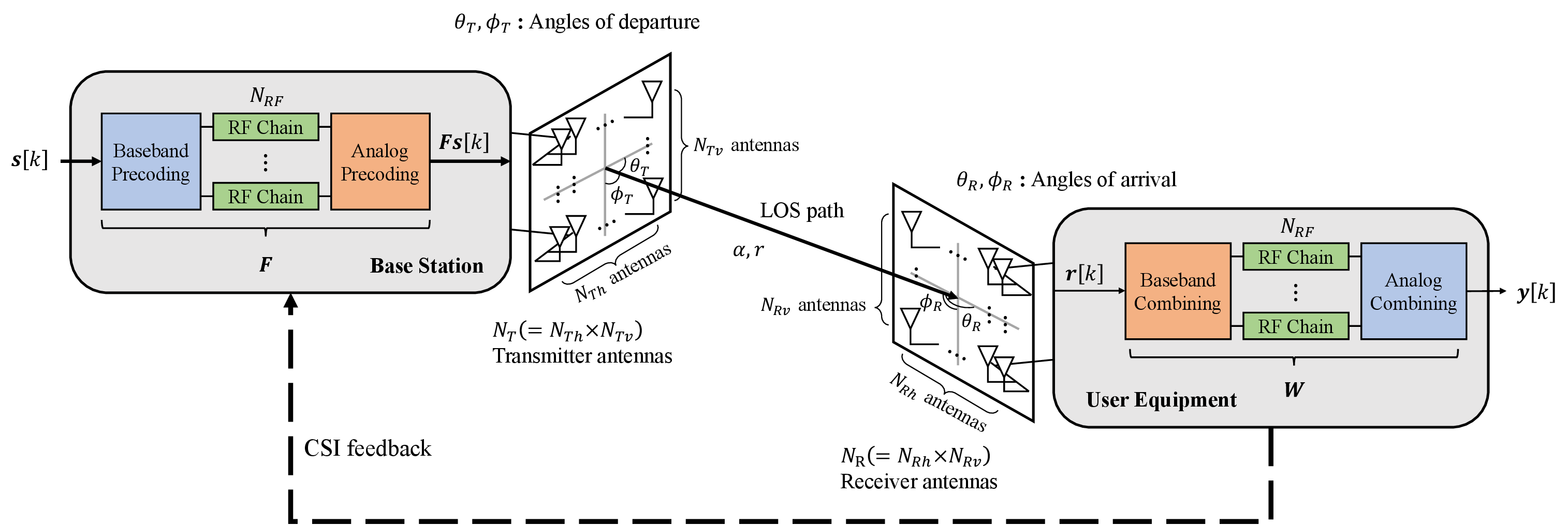}
	\caption{Illustration of THz MIMO systems with $N_T$ transmission antennas, $N_R$ receiver antennas, and $N_{RF}$ RF chains.}
	\label{fig:System_channel}
\end{figure*}

\section{THz MIMO System Model}
In this section, we discuss the near-field channel model for THz communications.
We also provide a brief discussion of the conventional channel estimation techniques.

\subsection{Downlink THz OFDM System Model}
\begin{figure*}[t]
	\centering
	\includegraphics[width=0.8\linewidth]{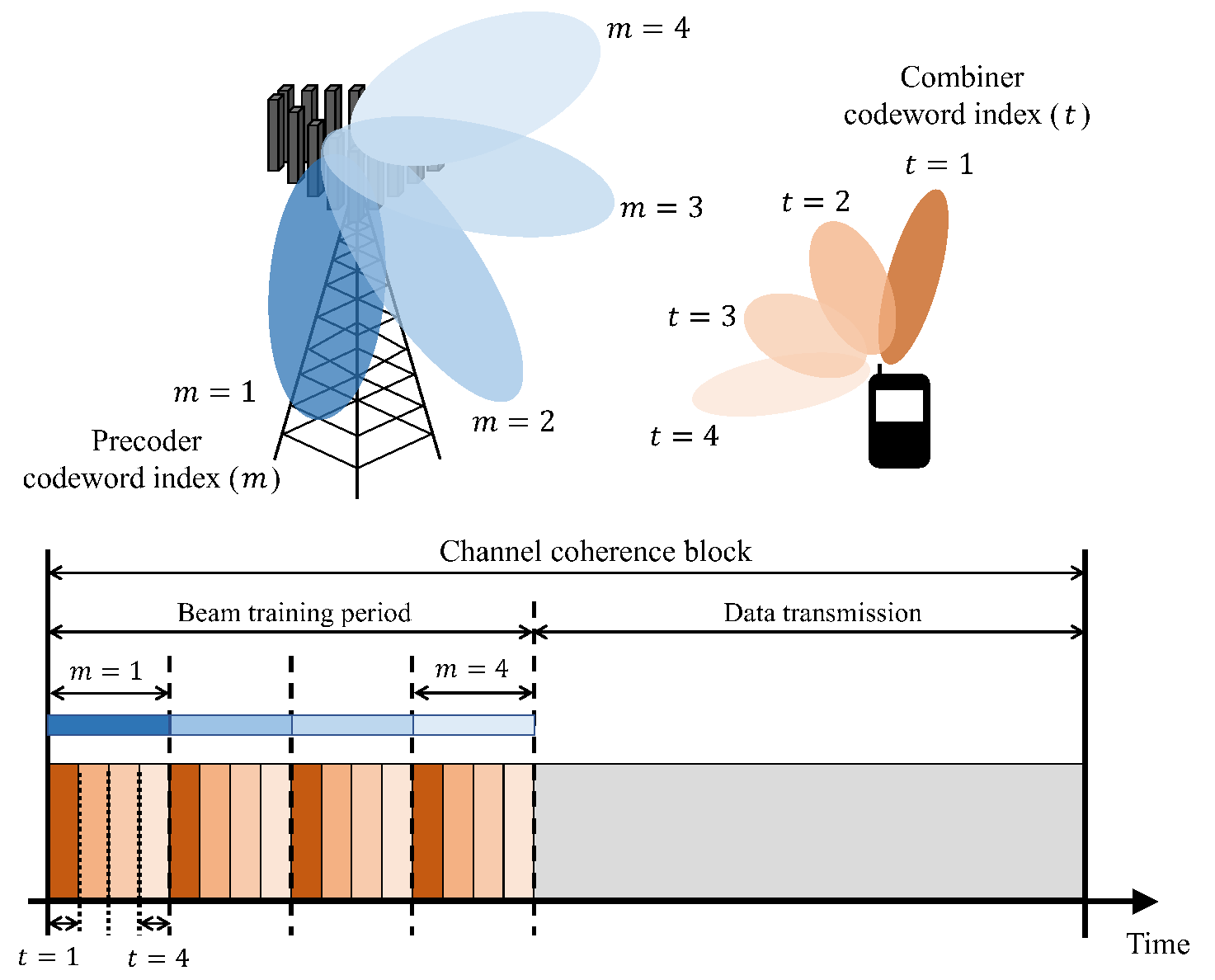}
	\caption{Example of the time scale of sub-frames and time frames in channel coherence block. In this illustration, we set $M$ time frames for precoding at the BS side and $T$ sub-frames for combining at the UE side. The UE receives a pilot signal for each transmitted precoder codeword by measuring all possible combiner codewords.}
	\label{fig:slot_time}
\end{figure*}

We consider a wideband downlink THz MIMO-OFDM systems where the BS equipped with $N_{T}=N_{Th} \times N_{Tv}$ planar antenna array serves the user equipment (UE) equipped with $N_{R}=N_{Rh} \times N_{Rv}$ planar antenna array (see Fig.~\ref{fig:System_channel}).
We assume that both BS and UE have $N_{RF}$ RF chains.
The number of OFDM subcarriers is $N_{s}$, among which $K$ subcarriers are used for the downlink pilot transmission.
Specifically, the downlink training period consists of $M$ successive time frames, each of which is divided into $T$ sub-frames.
In this work, we use $\mathcal{K}=\lbrace 1,\cdots,K\rbrace$, $\mathcal{M}=\lbrace 1,\cdots,M\rbrace$, and $\mathcal{T}=\lbrace 1,\cdots, T\rbrace$ to denote the sets of indices for pilot subcarriers, time frames, and sub-frames, respectively (see Fig.~\ref{fig:slot_time}).

In the THz MIMO system, the demodulation reference signal (DMRS) is transmitted after the beamforming to deal with the path loss and directivity problems of THz transmission.
In order to do so, the beam direction between the BS and the UE should be identified before the DMRS transmission.
In the 5G NR systems, the two-step beam management has been used to this end.
In the first stage called \textit{beam sweeping} stage, the BS transmits the synchronization signal block (SSB) beam codewords.
After receiving the SSB measurements, a mobile feeds back the index of a beam associated with the highest reference signal received power (RSRP).
For the area obtained from this process, the BS transmits multiple CSI-RS beam codewords to fine-tune the beam direction in the \textit{beam refinement} stage.

In the THz communication system, the received pilot signal (e.g., CSI-RS in 5G NR~\cite{3GPP}) $\mathbf{r}_{m}[k]\in\mathbb{C}^{N_{R}\times 1}$ of UE associated with $k$-th pilot subcarrier at $m$-th time frame is given by
\begin{equation}
\mathbf{r}_m[k] = \mathbf{H}[k]\mathbf{f}_{m}s_m[k] +\mathbf{v}_m[k],\quad m\in\mathcal{M},\, k\in\mathcal{K},
\end{equation}
where $\mathbf{H}[k]\in\mathbb{C}^{N_{R}\times N_{T}}$ is the downlink channel matrix, $\mathbf{f}_{m}\in\mathbb{C}^{N_{T}\times 1}$ is the beamforming vector, $s_{m}[k]\in\mathbb{C}$ is the pilot symbol, and $\mathbf{v}_{m}[k]\in\mathbb{C}^{N_{R}\times 1}$ is the additive Gaussian noise vector of $k$-th pilot subcarrier at $m$-th time frame.
During $T$ successive sub-frames, the UE sequentially employs combining vectors $\mathbf{w}_{1},\cdots,\mathbf{w}_{T}\in\mathbb{C}^{N_{R}\times 1}$ to obtain the processed signal $y_{m,k}[k]$:
\begin{equation}
y_{m,t}[k] = \mathbf{w}_{t}^{\textrm{H}}\mathbf{H}[k]\mathbf{f}_{m}s_{m}[k]+n_{m,t}[k],\quad t\in\mathcal{T},\, m\in\mathcal{M},\, k\in\mathcal{K}
\end{equation}
where $n_{m,t}[k]=\mathbf{w}_{t}^{\textrm{H}}\mathbf{v}_{m}[k]$.
By combining the processed signals into $M\times T$ pilot signal matrix $\mathbf{Y}[k]$, we have
\begin{equation}
\mathbf{Y}[k]=\mathbf{W}^{\textrm{H}}\mathbf{H}[k]\mathbf{F}\mathbf{S}[k]+\mathbf{N}[k],\quad k\in\mathcal{K},
\end{equation}
where $\mathbf{F}=[\mathbf{f}_{1},\cdots,\mathbf{f}_{M}]\in\mathbb{C}^{N_{T}\times M}$ is the beamforming matrix, $\mathbf{W}=[\mathbf{w}_{1},\cdots,\mathbf{w}_{T}]\in\mathbb{C}^{N_{R}\times T}$ is the combining matrix, and $\mathbf{S}[k]=\text{diag}(s_{1}[k],\cdots,s_{M}[k])$ is the pilot symbol matrix.
Finally, by vectorizing and concatenating the pilot signal matrices of $K$ pilot subcarriers, we obtain the measurement vector $\mathbf{y}$:
\begin{equation}
\mathbf{y}=[\text{vec}(\mathbf{Y}[1])^{\textrm{T}},\cdots,\text{vec}(\mathbf{Y}[K])^{\textrm{T}}]^{\textrm{T}}.
\end{equation}

\subsection{Near-field THz MIMO Channel Model}
In the THz communication systems, beamforming technique exploiting the massive MIMO antenna array is needed to compensate for the path loss.
In fact, main purpose of the beamforming technique is to concentrate the signal power on the narrow spatial region\footnote{Based on Friis' law, signal attenuation is proportional to the square of the carrier frequency.}.
One of the notable characteristics of the THz communications is that the Rayleigh distance, the boundary between the near-field and far-field, is fairly large.
The Rayleigh distance is formally defined as $Z=\frac{2D^2}{\lambda_c}$, where $D$ is the array aperture (i.e., maximum physical length of array antenna) and $\lambda_c$ is the wavelength of the carrier~\cite{selvan2017fraunhofer}.
Since the number of antenna arrays and the carrier frequency of the THz systems are far larger than those of the conventional mmWave systems, one can easily see that the Rayleigh distance $Z$ of THz systems is much longer than that of mmWave systems.
Note also that the electromagnetic signal in a near-field region is propagated in the form of a spherical wave so that the channel expressed by the far-field array response vector corresponding to the planar wavefront cannot reflect the THz channel properly.

\begin{figure*}[t]
	\centering
	\includegraphics[width=0.6\linewidth]{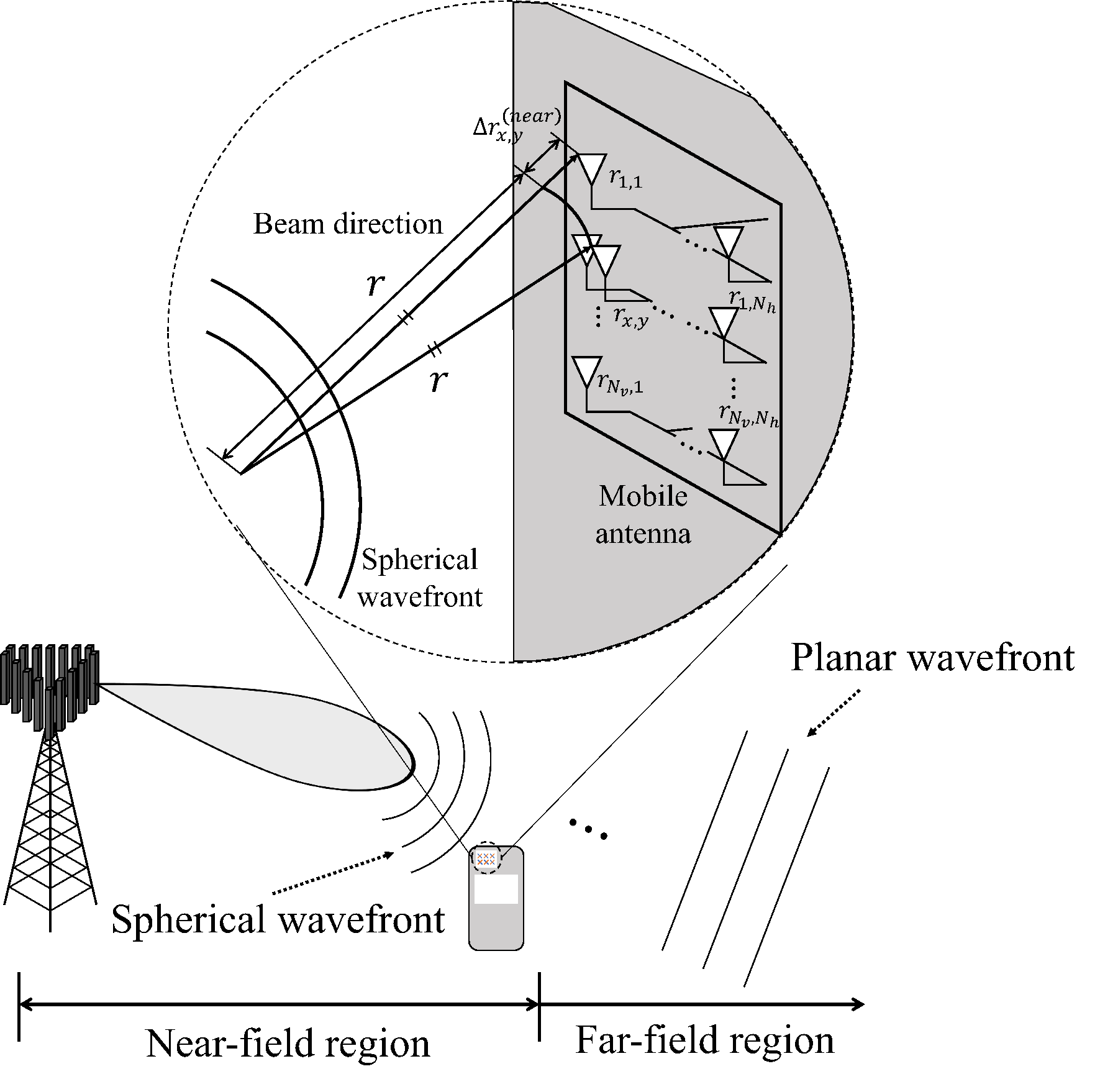}
	\caption{Illustration of near-field and far-field wavefront and the distance between$(0,0)$-th antenna and $(x,y)$-th antenna in planar array antenna.}
	\label{fig:near_field_channel}
\end{figure*}

To handle the problem, we need to carefully re-define the THz channel model in the near-field region.
Let $r_{x,y}$ be the distance from the $(x,y)$-th antenna of a planar array to the mobile, then the time delay between the signals transmitted from the $(x,y)$-th antenna and the $(1,1)$-th antenna (reference antenna) is $\frac{\Delta r_{x,y}}{c}$ where $\Delta r_{x,y}=r_{x,y} - r$ and $r=r_{1,1}$ is the distance between the reference antenna and the mobile.
Also, the corresponding phase shift in the frequency-domain is given by $e^{j2\pi f_c \frac{\Delta r_{x,y}}{c}} = e^{j\pi \frac{\Delta r_{x,y}}{d}}$ where $f_c$ is the subcarrier frequency, $d$ is the antenna spacing, $c=f_c\lambda_c$ and $d=\frac{\lambda_c}{2}$.

In the conventional far-field channel where the signals are transmitted in parallel to the antenna elements, $\Delta r_{x,y}$ is expressed as
\begin{align}
    \Delta r_{x,y}^{(\text{far})} &= -(x-1)d\cos\theta_{\text{az}}\sin\theta_{\text{el}} - (y-1)d\sin\theta_{\text{az}}\sin\theta_{\text{el}} \\
    &= -(x-1)d\theta - (y-1)d\phi,
\end{align}
where $\theta = \cos\theta_{\text{az}}\sin\theta_{\text{el}}$ and $\phi = \sin\theta_{\text{az}}\sin\theta_{\text{el}}$.
Then, the far-field planar array response vector can be expressed as a function of channel angle components $\theta$ and $\phi$:
\begin{align}
    \mathbf{a}_{\text{far}}(\theta,\phi) &= \Big[ e^{j\pi\frac{\Delta r_{1,1}^{(\text{far})}}{d}} , \cdots, e^{j\pi\frac{\Delta r_{N_x,N_y}^{(\text{far})}}{d}} \Big]^T \\
    &= [1, \cdots, e^{-j\pi((N_x-1)\theta + (N_y-1)\phi)}]^T\\
    &= \mathbf{a}_{\text{far},x}(\theta) \otimes \mathbf{a}_{\text{far},y}(\phi),
    \label{eq:far_field_a}
\end{align}
where $\mathbf{a}_{\text{far},x}(\theta) = [1, \cdots, e^{-j\pi (N_x - 1)\theta}]^T$ and $\mathbf{a}_{\text{far},y}(\phi) = [1, \cdots, e^{-j\pi (N_y-1)\phi}]^T$.

In the near-field channel, $\Delta r_{x,y}$ depends on both angle and distance due to the spherical radiation wavefront (see Fig.~\ref{fig:near_field_channel}).
To be specific, when we consider the spherical coordinate system where the position of reference antenna $r_{1,1}$ is set to $(0,0,0)$, then the coordinates of the $(x,y)$-th antenna and the mobile are $((x-1)d, (y-1)d, 0)$ and $(r\cos\theta_{\text{az}}\sin\theta_{\text{el}}, r\sin\theta_{\text{az}}\sin\theta_{\text{el}},$ $r\cos\theta_{\text{el}})$, respectively.
Thus, $\Delta r_{x,y}$ is given by
\begin{align}
    \nonumber \Delta r_{x,y}^{(\text{near})} &= \sqrt{\left( r\cos\theta_{\text{az}}\sin\theta_{\text{el}} - (x-1)d)^2 + (r\sin\theta_{\text{az}}\sin\theta_{\text{el}} - (y-1)d)^2 + (r\cos\theta_{\text{el}} \right)^2} - r \\
    &\approx -((x-1)d\cos\theta_{\text{az}}\sin\theta_{\text{el}} + (y-1)d\sin\theta_{\text{az}}\sin\theta_{\text{el}}) \label{eq:approx_r} \\
    \nonumber & +\frac{d^2}{2r}((x-1)^2 + (y-1)^2 - ((x-1)\cos\theta_{\text{az}}\sin\theta_{\text{el}} + (y-1)\sin\theta_{\text{az}}\sin\theta_{\text{el}})^2) \\
    &= -(\underbrace{(x-1)d\theta + (y-1)d\phi)}_{\Delta r_{x,y}^{\text{(far)}}} + \frac{d^2}{2r}((x-1)^2 + (y-1)^2 - ((x-1)\theta + (y-1)\phi)^2),
    \label{eq:near_r}
\end{align}
where the approximation in~\eqref{eq:approx_r} is obtained by the second order Taylor expansion (i.e., $\sqrt{1+x} \approx 1+\frac{x}{2} - \frac{x^2}{8}$).
Note, in contrast to the far-field array response model in~\eqref{eq:far_field_a}, the distance difference $\Delta r_{x,y}^{\text{(near)}}$ contains the distance $r$ between the UE and BS.
By exploiting~\eqref{eq:far_field_a} and~\eqref{eq:near_r}, the near-field array response vector of $(x,y)$-th antenna can be expressed as
\begin{align}
    \mathbf{a}_{\text{near}}(r,\theta,\phi) &= \Big[ e^{j\pi \frac{\Delta r_{1,1}^{(\text{near})}}{d}}, \cdots, e^{j\pi \frac{\Delta r_{N_x,N_y}^{(\text{near})}}{d}} \Big]^T \\
    &= \mathbf{D}(r,\theta,\phi) \mathbf{a}_{\text{far}}(\theta,\phi),
    \label{eq:a_near_steering}
\end{align}
where $\mathbf{D}(r,\theta,\phi) = \text{diag}(1,\cdots,e^{j\pi\frac{d}{2r}((N_x-1)^2+(N_y-1)^2-((N_x-1)\theta+(N_y-1)\phi)^2)})$ is the phase shift matrix induced by the spherical propagation of near-field signal.

In this work, we use the block-fading near-field LoS channel model where the downlink channel matrix $\mathbf{H}[k] \in \mathbb{C}^{N_T \times N_R}$ from the BS to the mobile for $k$-th subcarrier is expressed as\footnote{Note that in the THz communication systems, due to the significant path loss and directivity of THz band, the LoS component is dominant among the THz channel paths.}
\begin{align}
    \mathbf{H}[k] = \alpha e^{-j 2\pi \frac{r}{c} kf_s} \mathbf{a}_{\text{near}} (r,\theta_R,\phi_R) \mathbf{a}^{\textrm{H}}_{\text{near}} (r,\theta_T,\phi_T)
    \label{eq:near_field_channel}
\end{align}
where $\alpha$ is the complex path gain for path gain, $r$ is the distance between the BS and the mobile, $\theta_R$, $\phi_R$, $\theta_T$, and $\phi_T$ are channel angles of AoD and AoA, respectively.
Also, $c$ is light speed, and $f_s$ is subcarrier spacing.
One can see that $\mathbf{H}[k]$ is parametrized by a few channel parameters: AoAs $\{ \theta_R, \phi_R \}$, AoDs $\{ \theta_T, \phi_T \}$, distance $r$, and path gains $\alpha$.
Typically, the number of propagation paths is much smaller than the total number of antennas $N=N_{T}\times N_{R}$ (e.g., one or two propagation paths and $N=16\sim 256$) due to the high path loss and directivity of THz signal.
Since the THz channel is modeled by a small number of paths and each path consists of a few channel parameters, one can effectively estimate the channel by estimating sparse parameters $\lbrace \theta_R,\phi_R, \theta_T, \phi_T, r,\alpha \rbrace$ instead of the full-dimensional MIMO channel matrix $\mathbf{H}[k]$.

\subsection{Conventional THz MIMO Channel Estimation}
In the conventional THz MIMO channel estimation strategy, the full-dimensional channel matrix $\mathbf{H}[k]$ is estimated from $\mathbf{Y}[k]$ using the LS or LMMSE techniques.
Let $\tilde{\mathbf{y}}[k] = \text{vec}(\mathbf{Y}[k])$ and $\mathbf{h}[k] = \text{vec}(\mathbf{H}[k])$, then the LMMSE-based channel estimate is $\hat{\mathbf{h}}[k] = \text{Cov}(\mathbf{h}[k], \tilde{\mathbf{y}}[k]) \text{Cov} (\tilde{\mathbf{y}}[k],\\ \tilde{\mathbf{y}}[k])^{-1}\tilde{\mathbf{y}}[k]$.
To guarantee the accurate estimation of $\mathbf{h}[k]$, the number of measurements should be larger than or equal to the number of antenna elements.
However, in the practical THz systems, it is very difficult to acquire sufficient amount of measurements due to the large number of antennas.
For example, when the numbers of antennas at the BS and UE are $N_T = 32$ and $N_R = 8$, we need to allocate 2 subframes for the pilot transmission which occupies more than 15$\,$\% of a frame in 5G NR.

As an approach to address the problem, CS-based channel estimation technique has been widely used~\cite{choi2017compressed}.
In this approach, by mapping the complex gains to the sparse (path gain) vector, one can convert the channel estimation problem to the sparse recovery problem:
\begin{align}
\tilde{\mathbf{y}}[k]&=((\mathbf{F}\mathbf{S}[k])^{\textrm{T}}\otimes \mathbf{W}^{\textrm{H}})\mathbf{h}[k]+\text{vec}(\mathbf{N}[k])\\
&\approx((\mathbf{F}\mathbf{S}[k])^{\textrm{T}}\otimes\mathbf{W}^{\textrm{H}})\mathbf{A}\mathbf{g}[k]+\mathbf{n}[k], \\
&=\mathbf{\Phi}[k] \mathbf{g}[k] + \mathbf{n}[k], \label{eq:CSformulation}
\end{align}
where $\mathbf{\Phi}[k] = ((\mathbf{F}\mathbf{S}[k])^{\textrm{T}}\otimes\mathbf{W}^{\textrm{H}})\mathbf{A}$ is the sensing matrix, $\mathbf{A}=[\mathbf{a}_{\text{T}}^{\textrm{H}}(\bar{\phi}_{1})\otimes \mathbf{a}_{\text{R}}(\bar{\theta}_{1}),\cdots,\mathbf{a}_{\text{T}}^{\textrm{H}}(\bar{\phi}_{W})\otimes \mathbf{a}_{\text{R}}(\bar{\theta}_{W})]\in\mathbb{C}^{N_{T}N_{R}\times W}$ is the array steering matrix, $\mathbf{g}[k]\in\mathbb{C}^{W\times1}$ is the sparse path gain vector, and $\lbrace\bar{\theta}_{w},\bar{\phi}_{w}\rbrace_{w=1}^{W}$ is the set of quantized angles.

While the CS-based scheme is effective in dealing with the sparsity of the THz channel~\cite{choi2017compressed}, efficacy of the approach might be degraded in practical scenarios since there exists a mismatch between the true angle and the quantized angle.
One can try to reduce this mismatch by increasing the quantization level of an angle, but such trial will increase the column dimension of the sensing matrix $\mathbf{\Phi}[k]$, making the system model underdetermined.
In this underdetermined scenario where the number of unknown variables is far larger than the measurement size, the level of column correlation (e.g., mutual coherence of $\mathbf{\Phi}[k]$) will increase dramatically\footnote{The mutual coherence $\mu(\mathbf{\Phi})$ is defined as the largest magnitude of normalized inner product between two distinct columns of $\mathbf{\Phi}$: $$\mu(\mathbf{\Phi})=\max_{i\neq j} \frac{\vert \left< \mathbf{\Phi}_i, \mathbf{\Phi}_j \right> \vert}{\Vert \mathbf{\Phi}_i \Vert_2 \Vert \mathbf{\Phi}_j \Vert_2}$$
When the mutual coherence of two columns of $\mathbf{\Phi}$ is large and only one of these is associated with the nonzeros values in sparse vector $\mathbf{g}$, then it might not be easy to distinguish the right column (column associated with the nonzero value) from wrong one in the presence of noise.}, degrading the CS-based channel estimation quality severely.

\begin{figure}[t]
	\centering
	\includegraphics[width=0.8\linewidth]{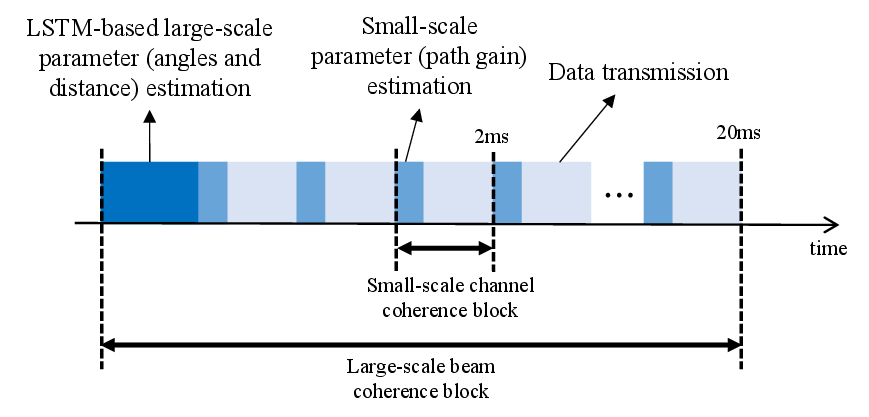}
	\caption{Example of time scale of channel parameters. In this example, coherence time interval of large-scale parameter (beam coherence interval) and small-scale parameter (channel coherence interval) is 20ms and 2ms, respectively.}
	\label{fig:time_scale}
\end{figure}

\section{Deep Learning-aided Channel Parameter Estimation Using Long Short-Term Memory}

Main idea of D-STiCE is to estimate the sparse channel parameters using the DL technique.
In our task, we exploit the fact that the THz channel is LoS dominant due to the severe signal attenuation in NLoS paths.
Therefore, what we need to do is to find a few parameters representing the LoS path in the spherical domain.
In order to facilitate the estimation of different types of channel parameters having distinct coherence time, we propose two step estimation process: 1) large-scale channel parameter extraction and 2) small-scale path gain estimation.
Large-scale parameters including the distance and angles are extracted via the LSTM network in a relatively long time interval (i.e., beam coherence interval)\footnote{In fact, the coherence time of angles and distance is about $40$ times larger than that of path gains~\cite{adhikary2017uplink}.}.
Since the (continuous) movement of a mobile has a good temporal correlation over time, the large-scale parameters can be readily inferred from the LSTM network.
By learning the temporal correlation between the 3D spherical coordinates (which are implicitly contained in the pilot measurements), D-STiCE can extract the large-scale near-field channel parameters accurately.
After identifying the large-scale parameters, we then update the small-scale (instantaneous) parameters during the beam coherence time.
The small-scale parameters are estimated via the conventional estimation technique (e.g., LS or LMMSE) in a channel coherence interval (see Fig.~\ref{fig:time_scale}).

\begin{figure}[t]
	\centering
	\includegraphics[width=0.9\linewidth]{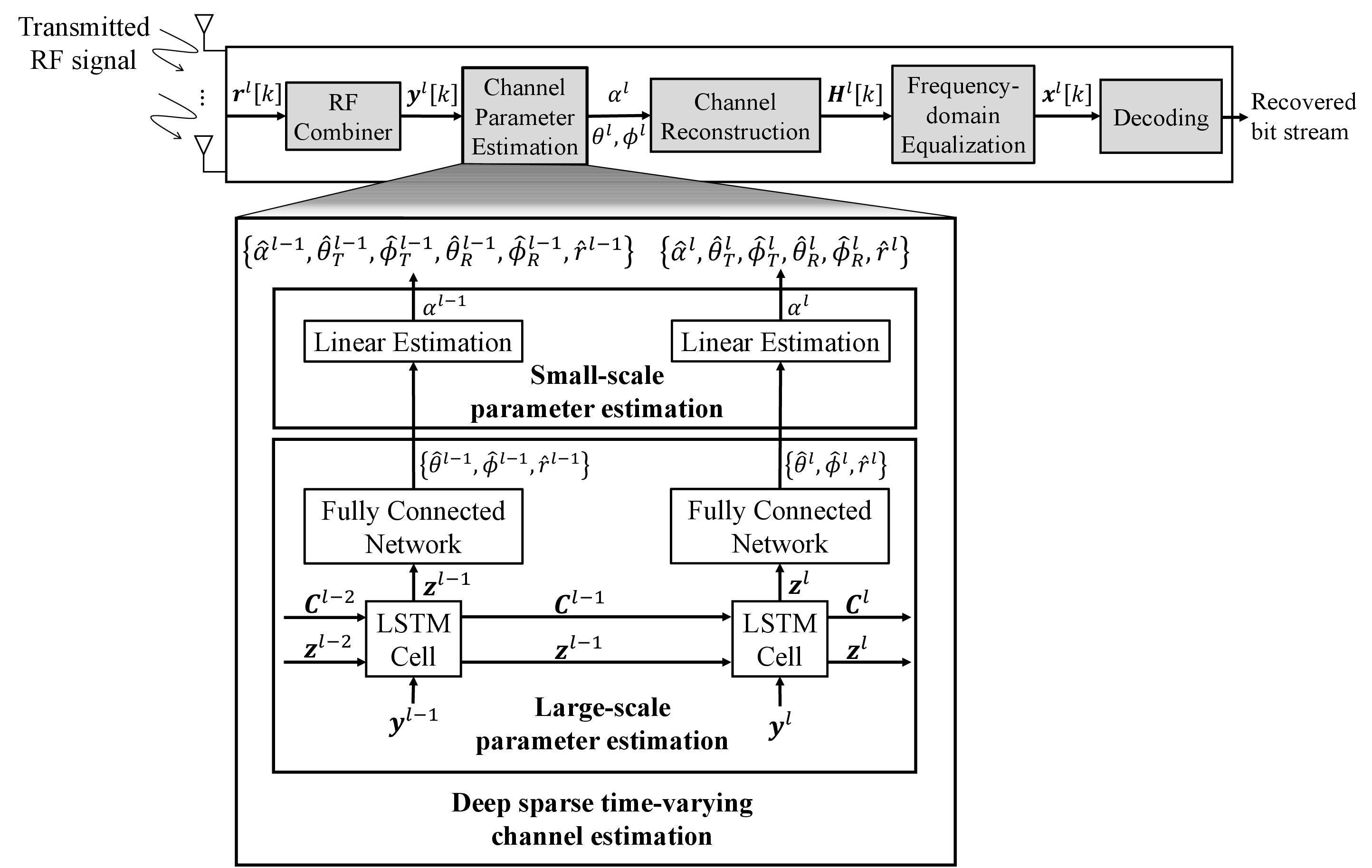}
	\caption{Overall receiver structure of the D-STiCE scheme.}
	\label{fig:network}
\end{figure}

\subsection{LSTM-based Large-scale Channel Parameter Estimation in Beam Coherence Time}
\begin{figure}[t]
	\centering
	\includegraphics[width=0.8\linewidth]{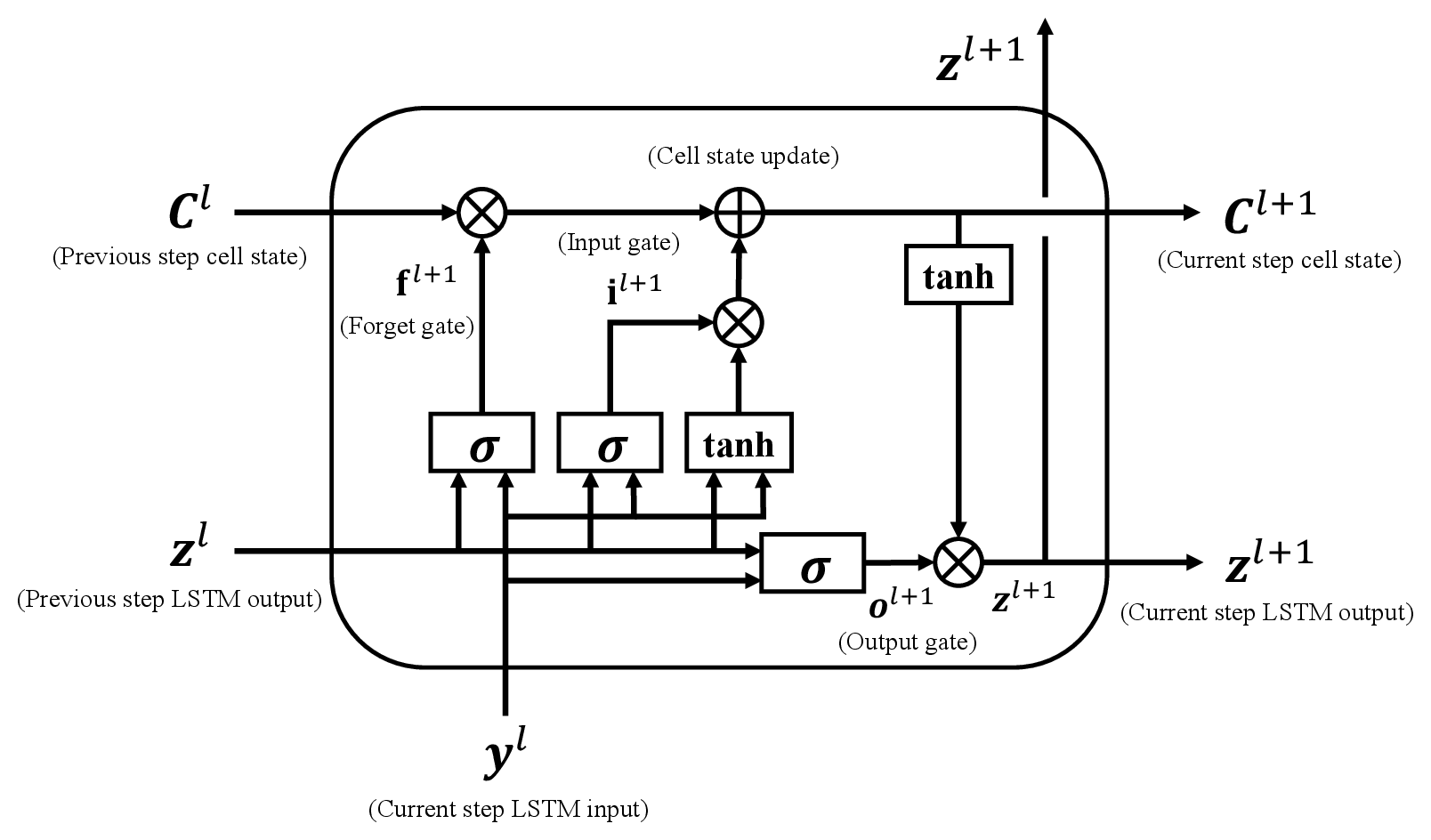}
	\caption{Detailed illustration of LSTM network}
	\label{fig:lstm_fig}
\end{figure}
In the large-scale channel parameter estimation, we exploit the combination of LSTM and fully connected (FC) networks to learn a complicated nonlinear mapping between the received downlink pilot signals ($\mathbf{y}^{1},\cdots,\mathbf{y}^{l}$) and the large-scale channel parameters ($\theta_R^l,\phi_R^l, \theta_T^l, \phi_T^l, r^l$):
\begin{equation}
\{\hat{\theta}_R^l, \hat{\phi}_R^l, \hat{\theta}_T^l, \hat{\phi}_T^l, \hat{r}^l\} = g(\mathbf{y}^1, \cdots, \mathbf{y}^l; \delta),
\end{equation}
where $\delta$ is the set of weights and biases.
The overall block diagram of the D-STiCE network is depicted in Fig.~\ref{fig:network}.

\subsubsection{LSTM Network}
The LSTM network consists of multiple LSTM cells, and each LSTM cell consists of \textit{cell state} and three gates, viz., \textit{input gate} $\mathbf{i}^l \in \mathbb{R}^{N_L \times 1}$, \textit{forget gate} $\mathbf{f}^l \in \mathbb{R}^{N_L \times 1}$, and \textit{output gate} $\mathbf{o}^l \in \mathbb{R}^{N_L \times 1}$ (see Fig.~\ref{fig:lstm_fig}).
Using the current input and the previous output, the forget, input, and output gates figure out the information to be removed, written, and read in the cell state.
Then, the updated cell state passes through the next LSTM cell sequentially~\cite{kim2021energy}.

The gating operations in the $l$-th LSTM cell are expressed as
\begin{align}
\mathbf{i}^{l} &= \sigma(\mathbf{W}_{i} \mathbf{y}^{l} + \mathbf{U}_{i} \mathbf{z}^{l-1} + \mathbf{b}_i) \label{eq:lstm1}\\
\mathbf{f}^{l} &= \sigma(\mathbf{W}_f \mathbf{y}^{l} + \mathbf{U}_{f} \mathbf{z}^{l-1} + \mathbf{b}_f) \\
\mathbf{o}^{l} &= \sigma(\mathbf{W}_o \mathbf{y}^{l} + \mathbf{U}_{o} \mathbf{z}^{l-1} + \mathbf{b}_o),
\end{align}
where $\sigma(x)=\frac{1}{1+e^{-x}}$ is the sigmoid function, $\mathbf{W}_i$, $\mathbf{W}_f$, $\mathbf{W}_o \in \mathbb{R}^{N_L \times KMT}$ are the weight matrices for the input, $\mathbf{U}_i$, $\mathbf{U}_f$, $\mathbf{U}_o \in \mathbb{R}^{N_L \times N_L}$ are the weight matrices for the previous output, and $\mathbf{b}_i$, $\mathbf{b}_f$, $\mathbf{b}_o \in \mathbb{R}^{N_L \times 1}$ are the bias vectors of input, forget, and output gate, respectively.
Then, the cell state is given by
\begin{align}
\mathbf{C}^{l} = \mathbf{f}^l \circ \mathbf{C}^{l-1} + \mathbf{i}^{l} \circ \text{tanh}(\mathbf{W}_c \mathbf{y}^{l} + \mathbf{U}^l_c \mathbf{z}^{l-1} + \mathbf{b}_c), \label{eq:lstm4}
\end{align}
where $\mathbf{W}_c \in \mathbb{R}^{N_L \times KMT}$ and $\mathbf{U}_c \in \mathbb{R}^{N_L \times N_L}$ are weight matrices, $\mathbf{b}_c \in \mathbb{R}^{N_L \times 1}$ is the bias, $\circ$ is the element-wise product, and $\text{tanh}$ is hyperbolic tangent function.
Finally, the output of LSTM network $\mathbf{z}^l \in \mathbb{R}^{N_L \times 1}$ is given by
\begin{align}
\mathbf{z}^l = \mathbf{o}^{l} \circ \text{tanh}(\mathbf{C}^{l}).
\label{eq:LSTM_output}
\end{align}

The forget, input, and output gates can be regarded as valves that regulate the delivery of cell state based on the temporal correlation $\rho$ between the past and current near-field array steering vectors (i.e., $\rho = \langle \mathbf{a}_{\text{near}}(r^{l-1}, \theta^{l-1}, \phi^{l-1}), \mathbf{a}_{\text{near}}(r^{l},$ $\theta^{l}, \phi^{l}) \rangle$).
For example, in scenarios where the mobility of a mobile device is low, temporal correlation of the large-scale parameters will be high (i.e., $\rho \approx 1$).
In this case, current channel parameters $\{\theta^{l}_R, \phi^{l}_R \}$ are strongly affected by the past channel parameters $\{\theta^{l-1}_R, \phi^{l-1}_R, \theta^{l-1}_T, \phi^{l-1}_T, r^{l-1}\}$, meaning that the forget gate vector $\mathbf{f}^l$ would be close to zero vector and the input gate would be close to one vector.
This helps the delivery of the previous cell state $\mathbf{c}^{l-1}$ containing the large-scale channel parameter information of past time slot (i.e., previous 3D spherical coordinates) to the current cell state $\mathbf{c}^{l}$.
Whereas, when the mobility of a mobile is high, the temporal correlation of the large-scale channel parameters is low (i.e., $\rho \approx 0$) so that it is desirable to focus only on the most recent pilot measurements.
In this case, learning process will enforce the forget and input gates such that these values will be close to one and zero, respectively.

We note that the similar effect can be obtained using \textit{Transformer}, an architecture specialized for extracting the correlation between inputs via attention mechanism.
Basically, Transformer and LSTM have been widely used to extract the temporal correlation of the input sequences (in our case, correlation of the geometric channel parameters over time).
While both techniques are similar in spirit, LSTM can be more effective in dealing with sequentially correlated inputs.
These days, Transformer is popularly used but it requires a large amount of memory and computationally intensive attention mechanism to handle both long and short-term correlations in tasks such as natural language processing (NLP) and speech recognition~\cite{vaswani2017attention, karita2019comparative}.
Since the LSTM network depends only on the output of the previous time step and the current measurement, the LSTM network can be implemented with much smaller memory space and simple hardware architecture.
Considering the memory overhead and stringent power constraint of the on-device AI as well as the design simplicity, LSTM can be a cost-effective choice.

\subsubsection{FC Network}
After the LSTM network, the FC network is employed to convert the extracted temporal channel features (LSTM output $\mathbf{z}^{l}$) into large-scale channel parameters $\{\theta^l_R, \phi^l_R,$ $\theta^l_T, \phi^l_T, r^l\}$.
The FC network consists of an input layer, hidden layers, and output layer.
The output $\mathbf{x}^0$ of the input layer is expressed as
\begin{align}
\mathbf{x}^{0} = \mathbf{W}^{0} \mathbf{z}^{l} +\mathbf{b}^{0}, \label{eq:initialFC}
\end{align}
where $\mathbf{z}^l$ is output vector of the LSTM network (see~(\ref{eq:LSTM_output})), $\mathbf{W}^{0} \in \mathbb{R}^{N_F \times N_L}$ is the weight matrix, and $\mathbf{b}^{0} \in \mathbb{R}^{N_F \times 1}$ is the bias vector of the input layer, respectively.
We use the batch normalization layer on $\mathbf{x}^0$ following the FC layer to obtain model parameters that are robust to signal distortion and noise.
Let $\mathbf{B}= [\mathbf{x}^{0,1},\cdots,\mathbf{x}^{0,d},\cdots,\mathbf{x}^{0,D}]^{\textrm{T}}$ be the stacked output vectors of the first FC layer, then the output vector $\widetilde{\mathbf{x}}^{0,d}$ of the batch normalization layer is~\cite{ioffe2015batch}
\begin{align}
\widetilde{x}^{0,d}_{i} = \gamma \Bigg( \frac{x^{0,d}_{i} - \mu_{\mathbf{B},i}}{\sqrt{\sigma^2_{\mathbf{B},i}}} \Bigg) + \beta, \label{eq:BN}
\end{align}
where $\mu_{\mathbf{B},i} = \frac{1}{D}\sum_{d=1}^D z^{l,d}_i$ and $\sigma^2_{\mathbf{B},i} = \frac{1}{D}\sum_{d=1}^D (z^{l,d}_i - \mu_{\mathbf{B},i})^2$ are mini-batch-wise mean and variance, respectively, $\gamma$ is the scaling parameter, $\beta$ is the shifting parameter, and $D$ is batch size.
As mentioned, the main purpose of using LSTM in D-STiCE is to learn a temporal variation of the large-scale channel parameters $(r,\theta,\phi)$ in the THz environments, not the random fluctuation in the complex gain $\alpha$.
To promote such behavior, we average out the small variations and distortions (e.g., Doppler shift, shadowing) in the pilot measurement $\mathbf{y}$ using the batch normalization process (see~\eqref{eq:BN}).

After the batch normalization, the output vector $\check{\mathbf{x}}^{0} = f_{\text{leaky-ReLU}} (\widetilde{\mathbf{x}}^{0})$ is generated by passing through the leaky rectified linear unit (leaky-ReLU) layer $f_{\text{leaky-ReLU}} = \max(0.1x, x)$.
Then, the output vector $\check{\mathbf{x}}^{0}$ passes through $N_m$ hidden layers consisting of the FC layer, batch normalization layer, and leaky-ReLU layer.
The output of the $i$-th hidden layer ($i=1,\cdots,N_m$) can be expressed as
\begin{align}
\check{\mathbf{x}}^{i} = f_{\text{leaky-ReLU}} \Bigg( \gamma^i \Bigg( \frac{\mathbf{W}^{i} \check{\mathbf{x}}^{i-1} + \mathbf{b}^{i} - \mathbf{\mu}}{\sqrt{\sigma^2}} \Bigg) + \beta^i \Bigg), \label{eq:leaky_ReLU}
\end{align}
where $\mathbf{W}^i \in \mathbb{R}^{N_F \times N_F}$ is the weight matrix and $\mathbf{b}^i \in \mathbb{R}^{N_F \times 1}$ is the bias vector of $i$-th hidden layer, respectively.
Finally, using the output vector $\check{\mathbf{x}}^{N_{m}}$ of the last hidden layer, we obtain the large-scale near-field channel parameter estimates $\{\theta^l_R, \phi^l_R, \theta^l_T, \phi^l_T, r^l\}$:
\begin{equation}
\{\hat{\theta}^l_R, \hat{\phi}^l_R, \hat{\theta}^l_T, \hat{\phi}^l_T, \hat{r}^l\} = \text{tanh}(\mathbf{W}^{\text{out}} \check{\mathbf{x}}^{N_M} +\mathbf{b}^{\text{out}}), \label{eq:FC_param}
\end{equation}
where $\mathbf{W}^{\text{out}} \in \mathbb{R}^{5 \times N_F}$ is weight vector, and $\mathbf{b}^{\text{out}} \in \mathbb{R}^{5 \times 1}$ is bias vector of output layer, respectively\footnote{Since the channel AoAs $\{ \theta_R, \phi_R \}$ and AoDs $\{ \theta_T, \phi_T \}$ are parameters of the periodic function, the output values should be limited in range to ensure one-to-one mapping. In D-STiCE, we use the hyperbolic tangent function to limit the output and scale the channel parameters in the range $(-1,1)$.}.

\subsection{Small-scale Channel Parameter Estimation in Channel Coherence Time}
Once the large-scale channel parameters are extracted, we estimate the small-scale channel parameters using a simple linear estimation technique.
After this, using both large-scale and small-scale channel parameters, we reconstruct the full-dimensional THz MIMO channel matrix $\mathbf{H}$, using which one can perform the channel equalization, log-likelihood ratio (LLR) generation, and packet decoding at the UE side (see Fig.~\ref{fig:network}).
While LSTM is effective in extracting the temporal correlation of the large-scale channel components, it might not be a good fit for estimating the small-scale channel parameters behaving like an independent and identically distributed (i.i.d.) random variable.

To be specific, after obtaining the large-scale channel parameters $\lbrace \hat{\theta}_R^l,\hat{\phi}_R^l, \hat{\theta}_T^l, \hat{\phi}_T^l, \hat{r}^l \rbrace$, we estimate the small-scale channel parameter using the conventional linear estimation technique.
Recall that the received pilot signal matrix of $k$-th pilot subcarrier is expressed as
\begin{equation}
\mathbf{Y}^l[k]=\mathbf{W}^{\textrm{H}}\mathbf{H}^l[k]\mathbf{F}\mathbf{S}[k]+\mathbf{N}^l[k],\quad \forall k\in\mathcal{K}.
\end{equation}
By vectorizing $\mathbf{Y}^l[k]$ into $\tilde{\mathbf{y}}^l[k]$, we have
\begin{align}
\nonumber \tilde{\mathbf{y}}^l[k]=\text{vec}(\mathbf{Y}^l[k])&=\alpha^l e^{-j2 \pi k f_s \frac{\hat{r}^l}{c}} (\mathbf{a}_{\text{near}}^{\textrm{H}}(\hat{r}^l,\hat{\theta}_T^l,\hat{\phi}^l_T)\mathbf{F}\mathbf{S}[k])^{\textrm{T}}\otimes \mathbf{W}^{\textrm{H}}\mathbf{a}_{\text{near}}(\hat{r}^l,\hat{\theta}_R^l,\hat{\phi}^l_R)+\text{vec}(\mathbf{N}^l[k])\\
&=\alpha^l e^{-j2 \pi k f_s \frac{\hat{r}^l}{c}} \mathbf{\Phi}[k] + \mathbf{n}^l[k],
\end{align}
where $\mathbf{\Phi}[k] = (\mathbf{a}_{\text{near}}^{\textrm{H}}(\hat{r},\hat{\theta}_T^l,\hat{\phi}^l_T)\mathbf{F}\mathbf{S}[k])^{\textrm{T}}\otimes \mathbf{W}^{\textrm{H}}\mathbf{a}_{\text{near}}(\hat{r},\hat{\theta}_R^l,\hat{\phi}^l_R)$ is the system matrix and $\mathbf{n}^l[k]$ is the vectorized noise vector.
Under this setting, the linear (e.g., LS and LMMSE) estimate of path gain vector $\alpha^l$ is
\begin{equation}
    \hat{\alpha}^l = e^{j2 \pi k f_s \frac{\hat{r}^l}{c}} (\boldsymbol{\Phi}[k]^{\textrm{H}}\boldsymbol{\Phi}[k])^{-1}\boldsymbol{\Phi}[k]^{\textrm{H}}\tilde{\mathbf{y}}^l[k].
\end{equation}

Finally, by substituting the acquired channel parameters $\lbrace \hat{\theta}_R^l,\hat{\phi}_R^l, \hat{\theta}_T^l, \hat{\phi}_T^l, \hat{r}^l, \hat{\alpha}^l \rbrace$ into the near-field THz MIMO channel model~\eqref{eq:near_field_channel}, we obtain the full-dimensional channel matrix $\widehat{\mathbf{H}}[k]$:
\begin{equation}
\widehat{\mathbf{H}}^l[k] = \hat{\alpha} e^{-j2 \pi k f_s \frac{\hat{r}}{c}} \mathbf{a}_{\text{near}} (\hat{r}^l,\hat{\theta}_R^l, \hat{\phi}_R^l) \mathbf{a}_{\text{near}}^{\textrm{H}} (\hat{r}^l, \hat{\theta}_T^l, \hat{\phi}_T^l).
\label{eq:est_channel_recon}
\end{equation}

\subsection{Loss Function Design and Training of D-STiCE}
In order to estimate the near-field channel parameters $\{ \hat{\theta}^l_R, \hat{\phi}^l_R, \hat{\theta}^l_T, \hat{\phi}^l_T, \hat{r}^l, \hat{\alpha}^l \}$, we design the deep neural network (DNN) such that the channel estimate $\widehat{\mathbf{H}}^l[k]$ in~\eqref{eq:est_channel_recon} is as close as possible to the true channel matrix $\mathbf{H}^l[k]$.
To this end, we use MSE-based loss function $J_{\delta}$ as
\begin{align}
J_{\delta} = \frac{1}{L} \sum_{l=1}^{L} \sum_{k=1}^{K}  \Vert \mathbf{H}^l[k] - \widehat{\mathbf{H}}^l[k; \delta] \Vert_F^2,
\label{eq:loss_function}
\end{align}
where $L$ and $K$ are the number of coherence intervals and the number of subcarriers, respectively.
Here, we express $\widehat{\mathbf{H}}^l[k]$ as $\widehat{\mathbf{H}}^l[k; \delta]$ to clearly identify that the channel estimate is parameterized by the DNN parameters $\delta$.
In our work, we obtain the ground-truth channel $\mathbf{H}^{l} [k]$ by collecting the channel samples generated from the realistic near-field THz MIMO downlink simulator (we will say more on this in Section V).
One might question that the synthetically generated channel might be different from the actual channel since the channel realizations can be changed by various factors such as temperature, humidity, and oxygen/foliage absorption.
Fortunately, we can circumvent this issue since the THz channel mainly consists of geometric parameters so that small variations can be readily considered as a small-scale parameter.

During the training phase, D-STiCE parameters $\mathbf{\delta}$ are updated using the loss function in~\eqref{eq:loss_function}.
Specifically, we update the parameter $\delta$ using the gradient descent method~\cite{ahn2021active}:
\begin{align}
\delta_{j} = \delta_{j-1} - \mu \sum_{l=1}^{L} \triangledown_{\delta} J_{\delta}^l,
\label{eq:BPTT}
\end{align}
where $j$ is training iteration, $\delta_j$ is the D-STiCE parameter of $j$-th iteration, and $\mu$ is the learning rate determining the amount of step to be updated in each iteration.
To guarantee the near-field THz channel estimation performance, we train D-STiCE until the absolute fractional difference of the loss $J_{\delta}$ is smaller than the predefined threshold (e.g., $\bigg \lvert \frac{J_{\delta_{j}} - J_{\delta_{j-1}}}{J_{\delta_{j-1}}} \bigg \rvert < \epsilon = 0.0001$).

\subsection{Complexity Analysis}
In this subsection, we briefly analyze the computational complexity of D-STiCE in the test phase.
In our analysis, we measure the complexity in terms of the number of floating point operations (flops).

Initially, in the LSTM cell, an input vector $\mathbf{y}^{l}$ is multiplied by the weights $\{ \mathbf{W}_{f}, \mathbf{W}_{i}, \mathbf{W}_{o}, \mathbf{W}_{c}\}$ (see~\eqref{eq:lstm1}-\eqref{eq:lstm4}).
Since the complex-valued matrix multiplication and bias addition operations are performed by dividing these operations in real and imaginary parts, the complexity of the aforementioned operations $\mathcal{C}_{cell_{1}}$ is $4(4KMT-1)N_{L} = 16KMTN_{L} - 4N_{L}$.
In a similar manner, the complexity $\mathcal{C}_{cell_{2}}$ of the matrix multiplications between $\mathbf{z}^{l-1} \in \mathbb{R}^{N_{L}}$ and $\{ \mathbf{U}_{f}, \mathbf{U}_{i}, \mathbf{U}_{o}, \mathbf{U}_{c}\} \in \mathbb{R}^{N_{L} \times N_{L}}$ can be expressed as
\begin{align}
\mathcal{C}_{cell_{2}} = 4(4N_{L}-1)N_{L} = 16N_{L}^{2} - 4N_{L}. \label{eq:complexity2}
\end{align}
Then, the bias additions ($N_{L}$ flops), sigmoid operations ($4N_{L}$ flops), tanh operations ($7N_{L}$ flops), element-wise multiplication ($N_{L}$ flops), and element-wise addition ($N_{L}$ flops) are performed 4 times, 3 times, 2 times, 3 times, and once, respectively (see~\eqref{eq:lstm1}-\eqref{eq:LSTM_output}).
Using~\eqref{eq:complexity2} and the operations we just mentioned, the complexity of the $L$ LSTM cells $\mathcal{C}_{LSTM}$ can be expressed as
\begin{align}
\mathcal{C}_{LSTM} &= L (\mathcal{C}_{cell_{1}} + \mathcal{C}_{cell_{2}} + 34N_{L}) \\
&= L (16KMTN_{L} + 16N_{L}^{2} - 8N_{L} + 34N_{L}) \\
&= 16LN_{L}^{2} + (16LKMT + 26L)N_{L}. \label{eq:complexity_lstm}
\end{align}

After passing through $L$ LSTM cells, the weight multiplication and bias addition are performed in the FC layer (see~\eqref{eq:initialFC}).
Since $\mathbf{W}^{0} \in \mathbb{R}^{N_{F} \times N_{L}}$ and $\mathbf{b}^{0} \in \mathbb{R}^{N_{F} \times 1}$, the complexity $\mathcal{C}_{in}$ of the initial FC layer is
\begin{align}
\mathcal{C}_{in} = (2N_{L}-1)N_{F} + N_{F} = 2N_{L} N_{F}.
\end{align}
Next, since the element-wise scalar multiplication ($N_{F}$ flops) and addition ($N_{F}$ flops) are performed twice in the batch normalization process before the leaky-ReLU function ($N_{F}$ flops), the complexity $\mathcal{C}_{in_{2}}$ of the corresponding operations is simply
\begin{align}
\mathcal{C}_{in_{2}} = 5N_{F}.
\end{align}

In the hidden layer, an input vector is multiplied by the weight $\mathbf{W}^{i} \in \mathbb{R}^{N_{F} \times N_{F}}$ and the bias $\mathbf{b}^{i} \in \mathbb{R}^{N_{F} \times 1}$ is added.
Then, the batch normalization ($4N_{F}$ flops) and leaky-ReLU operation ($N_{F}$ flops) are performed.
Therefore, the complexity $\mathcal{C}_{hide}$ of $N_{m}$ hidden layers can be expressed as
\begin{align}
\mathcal{C}_{hide} = N_{m} \cdot ((2N_{F}-1)N_{F} + N_{F} + 5N_{F}) = 2N_{m} N_{F}^{2} + 5N_{m} N_{F}.
\end{align}

Finally, in the last hidden layer, channel parameter estimates are extracted by multiplying the weight matrix $\mathbf{W}^{out} \in \mathbb{R}^{4P \times N_{F}}$ and then adding a bias vector $\mathbf{b}^{out} \in \mathbb{R}^{4P \times 1}$ to $\check{\mathbf{x}}^{N_{m}}$, where $P$ is the number of channel paths.
The complexity $\mathcal{C}_{out}$ of the last FC layer is
\begin{align}
\mathcal{C}_{out} = (2N_{F} -1)4P + 4P = 8PN_{F}.
\label{eq:complexity_out}
\end{align}
From~\eqref{eq:complexity_lstm} to~\eqref{eq:complexity_out}, the complexity $\mathcal{C}_{D-STiCE}$ of D-STiCE is summarized as
\begin{align}
\mathcal{C}_{D-STiCE} &= \mathcal{C}_{LSTM} + L \cdot(\mathcal{C}_{in} + \mathcal{C}_{in_{2}} + \mathcal{C}_{hide} + \mathcal{C}_{out}) \\
&= 16LN_{L}^{2} + 2N_{m} N_{F}^{2} + (16LKMT + 26L)N_{L} + (5N_{m}+2N_{L}+8P+5)N_{F}.
\end{align}

\begin{table*}[]
	\centering
	\caption{Comparison of Computational Complexity \\ $(K=64, L=10, N_{L}=500, N_{m}=2, N_{F}=500, P=3, W= N_T N_R, M = \frac{N_T}{2}, T= \frac{N_R}{2}, C_1 = 64, C_2= 3, C_3 = 10)$}
	\resizebox{0.99\columnwidth}{!}{
	\begin{tabular}{|c||c|c|c|c|}
		\hline
		\multirow{2}{*}{} & \multirow{2}{*}{the number of floating point operations (flops)} & \multicolumn{3}{c|}{Complexity for various $(N_T, N_R)$} \\ \cline{3-5} 
		& & $(16,4)$ & $(32,4)$ & $(32,8)$ \\ \hline
		\multirow{2}{*}{\textbf{D-STiCE}} & $16LN_{L}^{2} + 2N_{m} N_{F}^{2} + (16LKMT + 26L)N_{L}$ & \multirow{2}{*}{$3.34\times 10^{8}$}  & \multirow{2}{*}{$3.34\times 10^{8}$} & \multirow{2}{*}{$3.34\times 10^{8}$}       \\ 
		& $+ (5N_{m}+2N_{L}+8P+5)N_{F}$ & & & \\ \hline
		\multirow{2}{*}{\textbf{CS}}             &\multirow{2}{*}{$\left(2PW + \frac{P^4}{12} + \frac{5}{18}P^3 + \frac{47}{36} P^2 + P \right)KLMT$}                                      & \multirow{2}{*}{$4.96\times 10^{8}$}         & \multirow{2}{*}{$8.51\times 10^{8}$}       & \multirow{2}{*}{$1.30\times 10^{9}$}       \\ 
		& & & & \\	\hline
		\multirow{2}{*}{\textbf{CNN}}         &      \multirow{2}{*}{$(8C_{1} C_{2}^{2} + 2C_{1}^{2} C_{2}^{2} C_{3}) KL N_{R} N_{T}$}                                              & \multirow{2}{*}{$1.19\times 10^{9}$}       & \multirow{2}{*}{$3.11\times 10^{9}$}       & \multirow{2}{*}{$6.74\times 10^{9}$} \\ 
		& & & & \\ \hline
	\end{tabular}
	}
\end{table*}

In Table I, we compare the computational complexities of D-STiCE, CS-based channel estimation, and CNN-based channel estimation (see Appendix A for the detailed complexity derivation).
In order to examine the overall behavior, we compute the required flops for various numbers of antennas.
We observe that the complexity of D-STiCE is much smaller than that of conventional approaches.
For example, when $N_T = 32$ and $N_R = 4$, the complexity of D-STiCE is $50\%$ and $70\%$ lower than those of CS-based channel estimation and CNN-based channel estimation, respectively.
It is worth mentioning that the complexity of D-STiCE depends heavily on the dimension of the network parameters ($N_{L}$ and $N_{F}$), not the system parameters ($N_{T}$ and $N_{R}$).
For instance, when $N_{T}$ increases from 16 to 64, the number of flops of CNN-based channel estimation increases by 70\% but that for D-STiCE remains unchanged.
Thus, in the practical massive MIMO scenario where the numbers of transmit and receive antennas are large, the D-STiCE scheme still maintains a reasonable computational complexity.

\section{Practical Issues for D-STiCE Implementation}
In this section, we go over two practical issues in the D-STiCE implementation.
We first discuss the training data collection issue.
This issue is crucial since collecting a large amount of training data (i.e., THz channel) covering all the physical location ($r, \theta, \phi$) is very difficult in terms of energy and time-frequency resources.
We next discuss an environment compatibility issue.
If we re-train the D-STiCE whenever the wireless environment changes (e.g., indoor to outdoor and urban to rural), computational overhead and time for the re-training would be excessive, limiting the applicability of D-STiCE.

\subsection{Training Data Acquisition}
In order to acquire the optimal D-STiCE parameters $\mathbf{\delta}^{\ast}$, a large amount of training channel and pilot measurement samples are required.
However, in the THz MIMO system, it is very difficult to collect an abundant amount of \text{real} samples.
For instance, when we collect the channel measurements from the 5G NR system in a brute-force manner, it might take several days or weeks to collect measurements covering distinct scenarios (e.g., indoor/outdoor, urban/rural, and high speed/low speed).
Therefore, direct collection of real data might not be practical in terms of latency, resource utilization efficiency, and energy consumption.

As an answer to the problem at hand, we use a synthetically generated training dataset in this work.
Specifically, we generate the training data samples using the THz MIMO downlink simulator where mobile users are randomly distributed~\cite{3GPPchannel}.
By applying a realistic path loss model and geometric modeling for the synthetic data generation (we will say more on this in Section V), we can alleviate the mismatch between the real and synthetically generated datasets.

\begin{figure} 
	\begin{center}
		\includegraphics[width=0.9\linewidth]{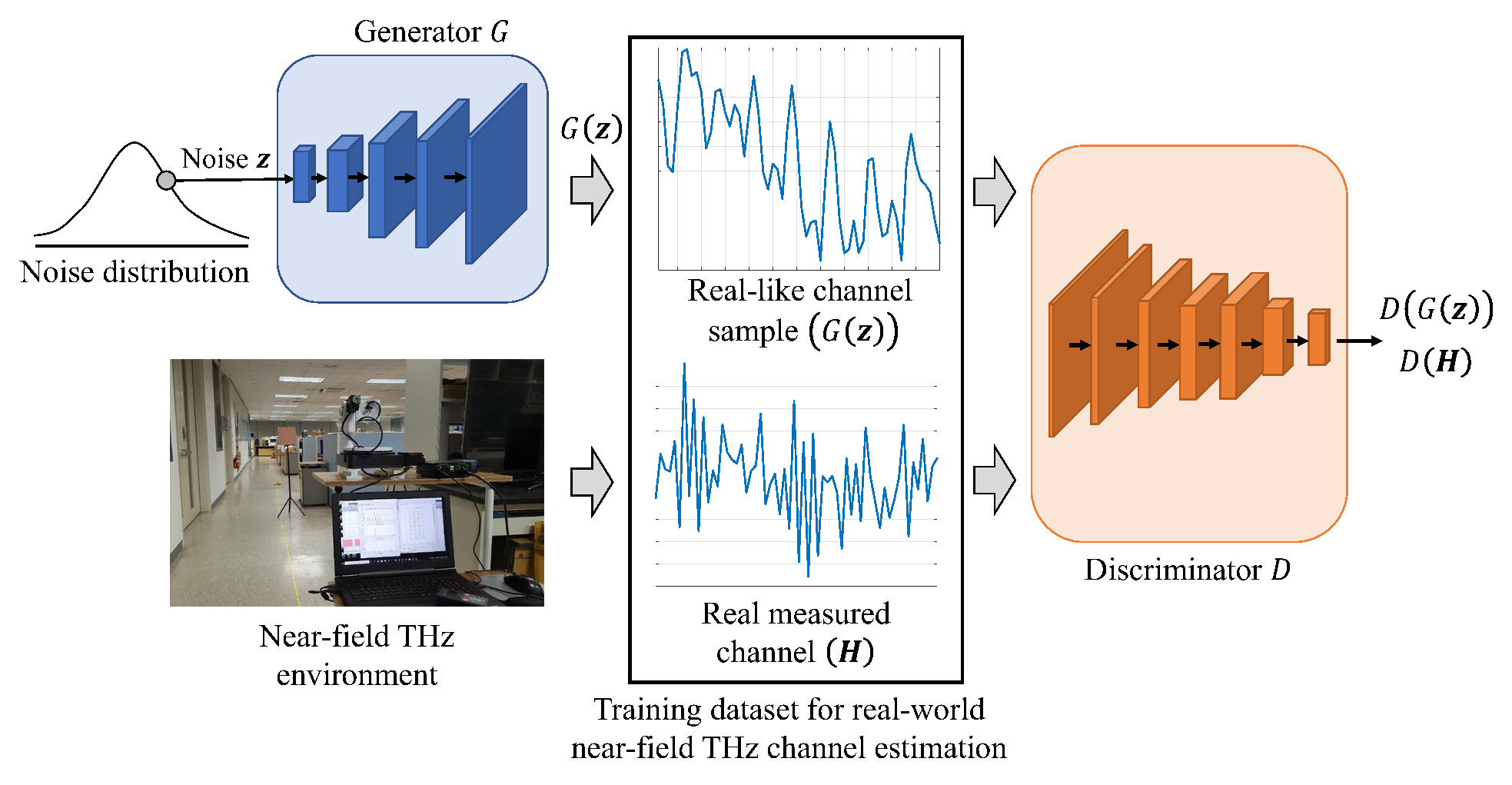}
	\end{center}
	\vspace{-2em}
	\caption{Illustration of GAN-based realistic channel data collection}
	\label{fig:near_GAN}
\end{figure}

One can also question that the use of synthetic data might not be sufficient to validate the direct applicability of D-STiCE in the real-world scenario.
For example, relying solely on synthetic data from system models or ray-tracing simulators can induce channel mismatch since manually designed simulators cannot completely reflect the details of the real-world (e.g., moving obstacles and material properties).
To address this potential problem, we exploit a generative adversarial network (GAN)-based dataset generation approach to generate massive real-like samples from the real ones~\cite{goodfellow2014generative}.
Note that GAN is a DL model that generates the samples approximating the input dataset (in our case, real measured THz MIMO channel matrix $\mathbf{H}$)~\cite{kim2023towards}.
The main components of GAN are a pair of DNNs, namely the \textit{generator} $G$ and the \textit{discriminator} $D$ (see Fig.~\ref{fig:near_GAN}).
In our case, $G$ tries to generate real-like near-field THz channel samples $G(\mathbf{z})$ from the random noise $\mathbf{z}$ and $D$ tries to distinguish whether the generated channel $G(\mathbf{z})$ is real or fake channel, respectively.
To train the GAN, we can adversarially train two DNNs using the min-max loss function, expressed as the cross-entropy\footnote{The cross-entropy between $x$ and $\hat{x}$ is defined as $H(x,\hat{x}) = - x \log(\hat{x}) - (1-x) \log(1- \hat{x})$.} between the distribution of generator output $G(\mathbf{z})$ and that of the real channel data $\mathbf{H}$:
\begin{align}
    \min_{G} \max_{D} \mathbb{E}_{\mathbf{H}}[\text{log}(D(\mathbf{H}))] + \mathbb{E}_{\mathbf{z}}[\text{log}(1-D(G(\mathbf{z})))],
\end{align}
where $D(\mathbf{H})$ is the discriminator output which corresponds to the probability of $\mathbf{H}$ being real (non-fake).
In doing so, when the training of GAN-based data generation is finished properly, the discriminator cannot evaluate whether the generated channel sample $G(\mathbf{z})$ is the generic or the synthetic (i.e., $D(G(\mathbf{z}))\approx 0.5$), which means that the generated THz channel sample is reliable and thus one can readily use the generated channel matrix $\mathbf{H}$ for the training of D-STiCE.
In particular, when compared to the manually designed channel models, GAN-based channel samples well represent complicated channel characteristics (e.g., foliage loss, atmospheric absorption, and cluttering density), so that GAN-based samples can be readily used for the D-STiCE training.

\begin{figure}
    \centering
    \includegraphics[width=0.7\linewidth]{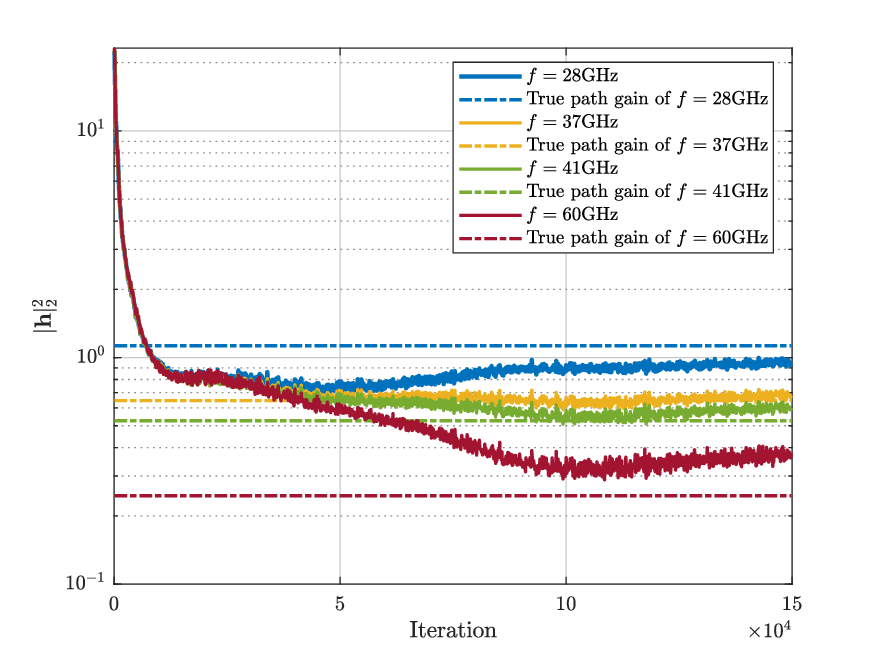}
    \caption{Evaluations of channel path gain as a function of GAN training iterations. We generate the channel data for $f=28,37,41$, and $60\,$GHz in every 100 iterations.}
    \label{fig:enter-label}
\end{figure}

In order to briefly show the similarity between the GAN-based dataset and real one, we measure the path gains of the GAN-based channel samples and those of the real channel samples as a function of the training iteration (see Fig.~\ref{fig:enter-label}).
In our numerical evaluations, we generate the narrowband geometric channel samples for the various center frequencies including $f=28,37,41$, and $60\,$GHz.
Specifically, the propagation channel model $\mathbf{h}\in \mathbb{C}^{N_{t} \times 1}$ is expressed as
\begin{align}
    \label{eq:channel_model} \mathbf{h} &= \sqrt{\frac{N_{t}}{L}} \sum_{l=1}^{L} \rho_{l} \mathbf{a}(\theta_{l}), \\
    \rho_{l} &\sim \mathcal{CN}(0,C) = \mathcal{CN}(0,\frac{P_{0}}{f^{2} R^{2}}), \label{eq:7}\\
    \mathbf{a}(\theta_{l}) &= \frac{1}{N_{t}} \left[ 1, e^{j\theta_{l}},\cdots,e^{j (N_{t}-1)\theta_{l}} \right]^{T}, \label{eq:8}
\end{align}
where $\rho_{l}$, $C$, $P_{0}$, $f$, $R$, $\theta_{l}$, $\mathbf{a}$, and $L$ are the complex gain, distance-dependent path loss, power gain, center frequency, distance, AoD, transmit array response vector associated with the $l$-th propagation path, and the number of paths, respectively.
One can readily observe from~\eqref{eq:7} that the channels at different center frequencies follow distinct path gain distributions.
As shown in Fig.~\ref{fig:enter-label}, we observe that the path gains of the GAN-based channels converge to those of the model-based channels for all center frequencies $f = 28, 37, 41,$ and $60$ GHz as the training iteration increases.
For example, the average path gain of the 28$\,$GHz channel data becomes 0.98 after 100,000 iterations, which is very close to the path gain of the benchmark, 1.1274.

\subsection{Environment Compatibility Issue}
Yet another practical issue is the environment compatibility. 
Due to the high penetration loss and movement of mobile devices or obstacles in the THz propagation environment, wireless channel characteristics between the BS and UE can be changed occasionally.
For example, when the UE moves from indoor to outdoor, distribution of the communication distance $r$ (e.g., 5$\,$m in the indoor room and 20$\,$m in the outdoor on average) and blockage ratio changes considerably so that the D-STiCE needs to be re-trained to deal with the changed scenario.
This issue is crucial for D-STiCE since the re-training in each environment will not be handy and incur the waste of energy, resources, and time.

To address the issue, we exploit the meta-learning, a special DL technique to solve new task with only a small number of training samples (in our case, pilot measurements)~\cite{kim2022massive}.
Essence of the meta-learning is to acquire the pre-trained model and then quickly train the desired function with a few training samples~\cite{finn2017model}.
So, as long as we have a pre-trained model which already learned the common channel characteristics (e.g., parametric sparsity in the 3D spherical domain, temporal correlation of $(r,\theta,\phi)$, LoS-dominant property), we can significantly reduce the required number of THz near-field channel samples.
Since what should be done for the new wireless environment is to learn the distinct channel features (e.g., distance and angle change) describing a new environment (this process is often called \textit{fine-tuning}), we can save the effort to re-train the whole model parameters $\delta$.

Basically, the meta-learning process for D-STiCE consists of two main steps: 1) temporal update of the D-STiCE parameters for given datasets and 2) identification of the centroid of the temporarily updated D-STiCE parameters.
In each iteration, we first temporarily update the D-STiCE parameters $\{ \widetilde{\delta}_{D_{i},t}\}_{i=1}^{M}$ for each given $M$ datasets $\{D_i\}_{i=1, \cdots, M}$.
To update the pre-trained model parameter $\delta$, we evaluate the loss of each dataset $J_{\widetilde{\delta}_{D_i,t}}^{D_i}$ with respect to temporally updated D-STiCE parameters.
We then update the D-STiCE parameters $\delta$ in a direction to minimize the sum of losses $\sum_{i=1}^{M} J_{\widetilde{\delta}_{D_i,t}}^{D_i}$.
In doing so, we obtain the pre-trained model of D-STiCE that learns the common channel characteristics in the $M$ datasets $\{D_i\}_{i=1}^{M}$.
In the fine-tuning phase, we can use $\delta$ as an initialization point of D-STiCE, using which one can quickly learn the parametric near-field channel estimation in a new THz dataset $D_{M+1}$.

\section{Simulations and Discussions}
\subsection{Simulation Setup}
\begin{figure} 
	\begin{center}
		\includegraphics[width=0.6\linewidth]{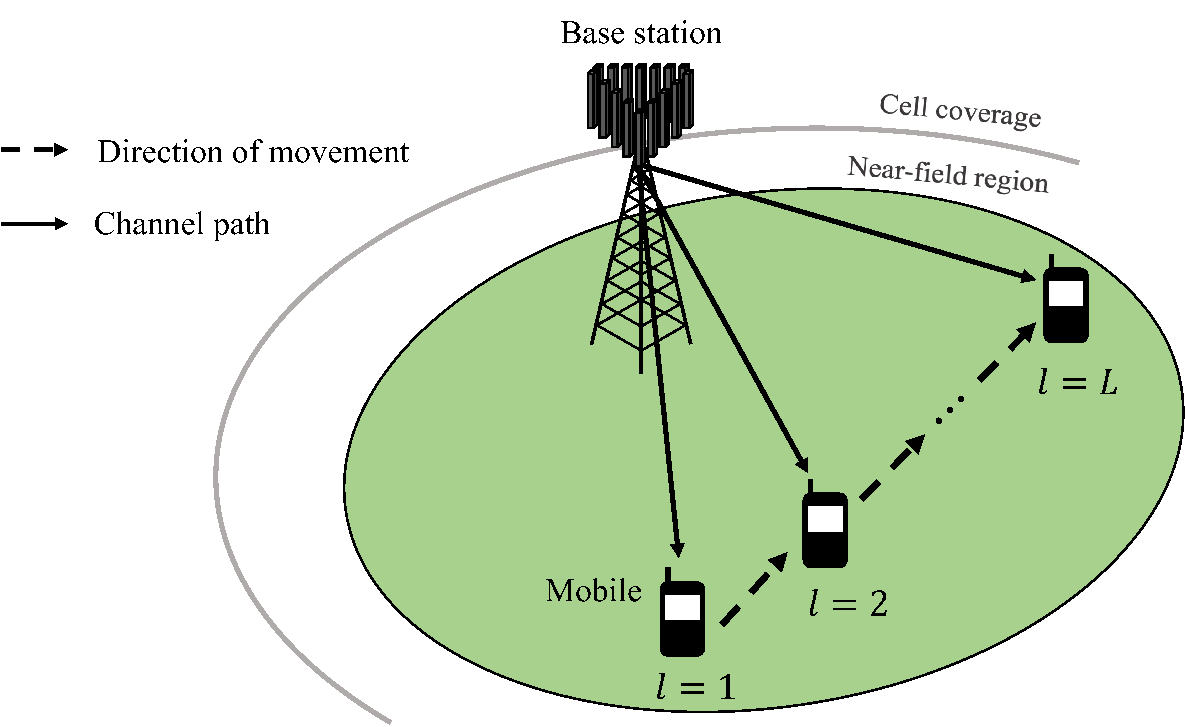}
	\end{center}
	\caption{Illustration of simulation environment. In our simulation, the UE are distributed within 10 m around the BS. The mobile user moves along the line trajectory with a constant speed of $3$ km/h.}
	\label{fig:simul_setup}
\end{figure}
In this section, we investigate the performance of D-STiCE.
In our simulations, we consider the near-field THz MIMO systems with $1\,$THz frequency band where the BS and UE are equipped with $N_T=64$ and $N_R=16$ antennas, respectively.
We set the number of RF chains to $N_{RF}=4$ and the antenna spacing to $d=\frac{\lambda}{2}$ for both the BS and UE.
Following 5G NR standard, we set the subcarrier spacing to $f_s = 120$ KHz~\cite{3GPP}.
We also set the numbers of pilot subcarriers, time frames, and sub-frames to $K=4$, $M=16$, and $T=16$, in each channel coherence interval\footnote{Note that our simulation parameters are designed by considering both the channel coherence time and beam training time.
In our scenario where the user is moving at 3$\,$km/h in the 1$\,$THz frequency band, the channel coherence time is approximately given by 152.3$\,\mu$s.
Since the symbol duration under the setting of 120$\,$kHz subcarrier spacing is approximately 9$\,\mu$s, one can transmit 16 symbols (approximately 144$\,\mu$s) for the downlink beam training within one channel coherence time.}.
The number of large-scale beam coherence intervals is $L=5$.
In our simulations, we use the beamforming matrix $\mathbf{F}$ and the combiner matrix $\mathbf{W}$ consisting of steering vectors that equally space the angle in the interval $(-\pi /2 , \pi /2]$.
To model the channel variation over the time, the UE moves along the line trajectory with a constant speed ($3$ km/h).
We assume that the small-scale channel parameter $\alpha$ is independently and identically distributed (i.i.d.) complex Gaussian random variable.

In the simulations of D-STiCE, we generate 200$\,$K, 10$\,$K, and $10\,$K samples for the training, validation, and testing, respectively.
The channel samples are generated using~\eqref{eq:near_field_channel}.
In the training process, we use an Adam optimizer with learning rate $10^{-4}$.
As performance metrics, we use the NMSE$=\Vert \mathbf{H} - \widehat{\mathbf{H}} \Vert^2_F / \Vert \mathbf{H} \Vert^2_F$ and BER.
For comparison, we use five benchmark schemes: 1) LS estimator, 2) LMMSE estimator, 3) CS-based channel estimation~\cite{cui2021near}, 4) CNN-based channel estimation~\cite{dong2019deep}, and 5) Transformer-based channel estimation.
To ensure the fairness of the computational complexity, we set the similar level of computational overhead for the Transformer-based channel estimation and proposed D-STiCE.
Specifically, we use 5 cell states and 2 vertical hidden layers for the LSTM network and 2 hidden layers for the FC network.
The number of hidden layer units in LSTM cell and FC network $N_L$ and $N_F$ are set to 576.
Also, we use 5 blocks of Transformer encoder consisting of 16 multi-heads and 1,216 dimensions for embedding.

\begin{figure}
	\begin{center}
		\includegraphics[width=0.72\linewidth]{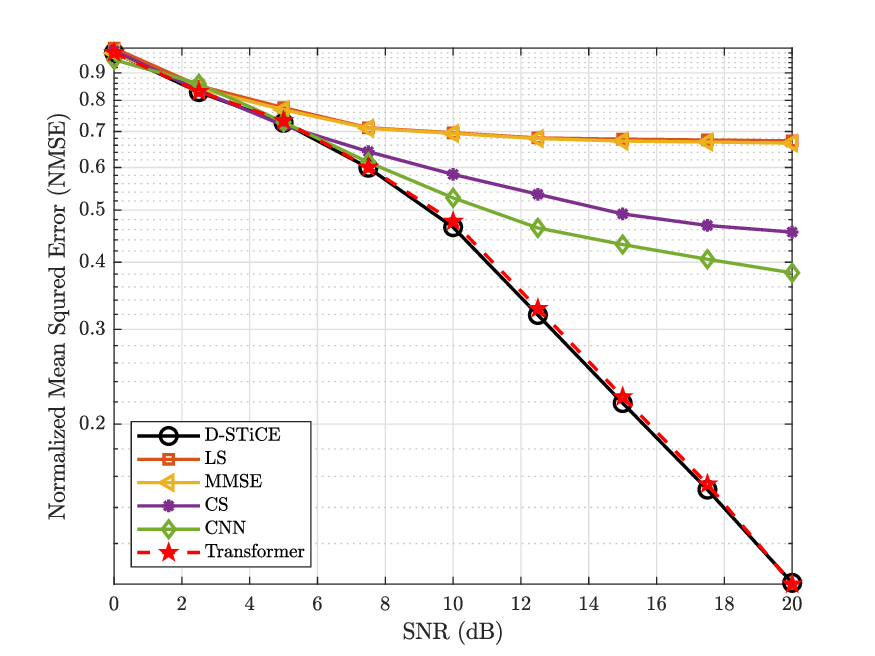}
	\end{center}
	\caption{Normalized MSE performance of channel estimation techniques as function of SNR.}
	\label{fig:sim_1}
\end{figure}

\subsection{Simulation Results}
In Fig.~\ref{fig:sim_1}, we plot the NMSE of D-STiCE and competing schemes as the function of SNR.
We observe that D-STiCE outperforms the conventional channel estimation algorithms by a large margin, especially in high SNR regime.
For example, when SNR $=20\,$dB, the NMSE of D-STiCE is less than $0.2$ while those of the conventional linear estimator (i.e., LS, LMMSE estimator), CS-based estimator, and the CNN-based estimator are $0.7$, $0.5$, and $0.4$, respectively.
While the trained D-STiCE exploits the sparsity of the spherical domain channel and thus it can accurately estimate the channel in the limited pilot scenario, conventional LS and MMSE techniques have no such mechanism so that the performance of these schemes is not that appealing.
We show that the performance of D-STiCE is even better than that of the CNN-based approach in the high SNR regime since D-STiCE can effectively leverage the temporal correlation between the past and current channel parameters to narrow down the range of the angle (and distance) search.
One can also see that the performance of D-STiCE and Transformer-based channel estimation is more or less similar.
Since the large-scale channel parameters are highly correlated to the adjacent measurements rather than the old ones, there should be no fundamental difference between two.

\begin{figure}
	\begin{center}
		\includegraphics[width=0.72\linewidth]{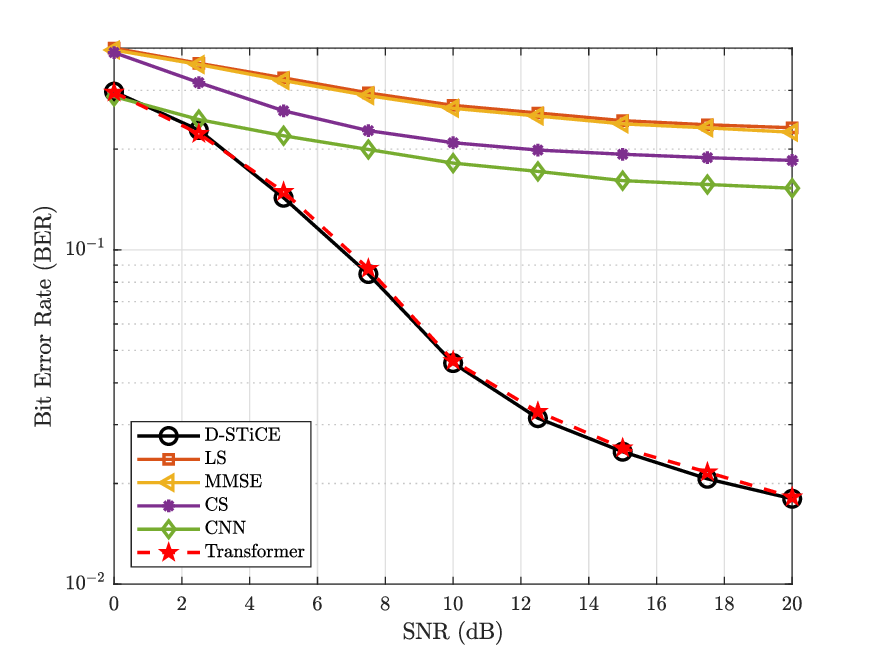}
	\end{center}
	\caption{BER performance of channel estimation techniques as function of SNR.}
	\label{fig:sim_2}
\end{figure}

To measure the impact of D-STiCE on the overall system performance, we next evaluate the BER performance.
Since our purpose is to check the channel estimation quality, we use a simple setup to detect the QPSK symbols using hard decision decoding after the channel estimation.
As shown in Fig.~\ref{fig:sim_2}, D-STiCE achieves more than $3\,$dB gain for all SNR regimes.
For instance, when compared to the BER performance at SNR $=20\,$dB, D-STiCE achieves more than one order of magnitude improvement in BER over the conventional channel estimation techniques.

\begin{figure}
	\begin{center}
		\includegraphics[width=0.72\linewidth]{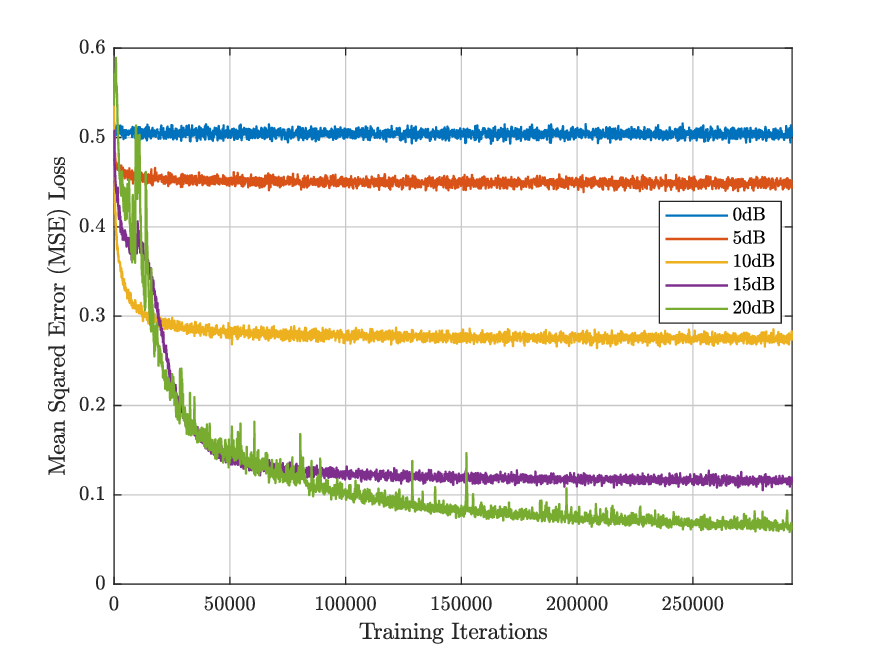}
	\end{center}
	\caption{Training loss of D-STiCE as a function of training iterations. We evaluate the MSE loss every 100 iterations.}
	\label{fig:sim_3}
\end{figure}

In Fig.~\ref{fig:sim_3}, we evaluate the MSE loss of D-STiCE as a function of the training iterations.
From the simulation results, we observe that the MSE loss of D-STiCE decreases gradually and finally converges.
For example, the MSE loss converges to 0.12 at the SNR $=15\,$dB after 100,000 training iterations.
We also observe that the MSE loss is further reduced when the SNR increases.
For instance, the MSE loss converges to 0.5 in $0\,$dB and while that in $15\,$dB converges to 0.12.
We note that this behavior is reasonable since the accurate estimation of large-scale parameters will do good for the small-scale parameter estimation.

\begin{figure}
	\begin{center}
		\includegraphics[width=0.72\linewidth]{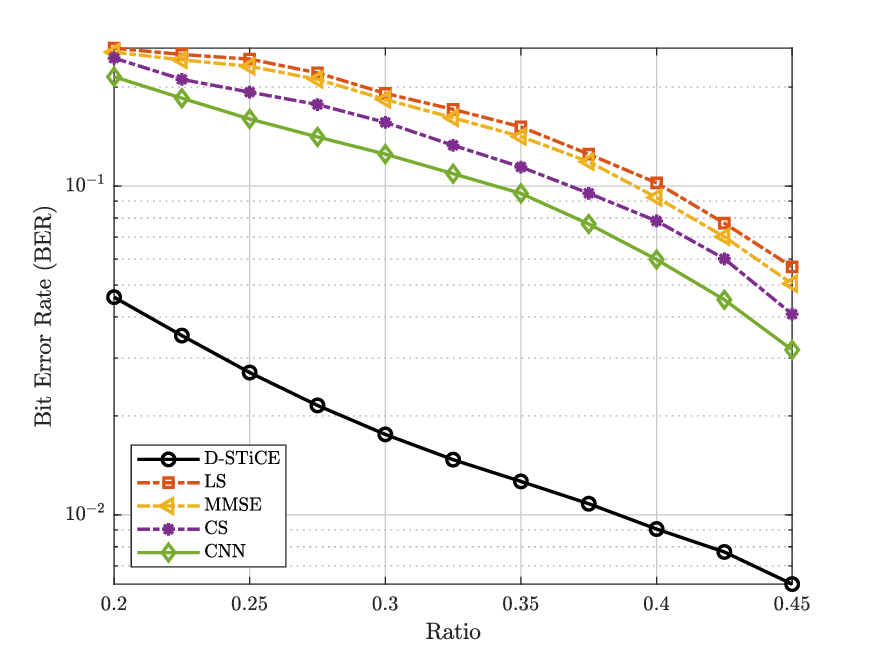}
	\end{center}
	\caption{BER performance of channel estimation schemes as function of ratio between received pilot dimension and channel dimension. In this simulation, we fix the SNR as 15dB.}
	\label{fig:sim_4}
\end{figure}
In Fig.~\ref{fig:sim_4}, we plot the BER performance as a function of the ratio between the number of received pilot signals and full-channel coefficients (i.e., $\text{ratio}=\frac{MT}{N_R N_T}$).
From the simulation results, we observe that D-STiCE outperforms the conventional channel estimation schemes by a large margin even when the number of pilot measurements is less than 0.25 of the full-channel coefficients.
For example, when compared to the linear, CS-based, and CNN-based channel estimators, D-STiCE requires less than half of the pilot signals at $\text{BER} = 0.05$ because the D-STiCE just requires a few sparse parameters $\lbrace \hat{\theta}_R^l,\hat{\phi}_R^l, \hat{\theta}_T^l, \hat{\phi}_T^l, \hat{r}^l, \hat{\alpha}^l \rbrace$ to reconstruct the full-channel (see Section III.A).
Since there is no such mechanism for the conventional channel estimation techniques, the conventional schemes perform well only when the number of pilots is large enough.

\section{Conclusion}
In this paper, we proposed a DL-based parametric channel estimation technique for the near-field THz MIMO systems.
Using the property that the near-field THz channel can be expressed as a few channel parameters in the spherical domain, viz., AoDs, AoAs, distance, and path gain, the proposed D-STiCE estimates the large-scale sparse channel parameters via LSTM-based DNN.
Then, by combining the large-scale parameters (angles and distance) and small-scale path gain parameter, we could reconstruct the THz MIMO channel matrix.
From the numerical evaluations in the 6G THz environment, we verified that the proposed D-STiCE is effective in estimating the near-field THz downlink environments.
Given the trend of AI technology being actively applied to wireless communication, we believe that D-STiCE will serve as an effective channel estimation framework in the upcoming 6G systems.

\begin{appendices}
\section{Proof of the computational complexities in Table I}
In this appendix, we analyze the computational complexities of CS-based channel estimation and CNN-based channel estimation in Table I.
In the OMP algorithm, a a submatrix $\mathbf{\Phi}_{l}[k]$ of $\mathbf{\Phi}[k]$ having the maximum correlation between the residual vector $\mathbf{r}^{j-1}$ is chosen, and $\mathbf{g}^{j}[k]$ is estimated (i.e., $\hat{\mathbf{g}}^{j}[k] = \left( \mathbf{\Phi}[k]_{\Omega_{j}}^{\textrm{H}} \mathbf{\Phi}[k]_{\Omega_{j}}\right)^{-1} \mathbf{\Phi}[k]_{\Omega_{j}}^{\textrm{H}} \tilde{\mathbf{y}}[k]$).
Using the Cholesky decomposition, we the complexity of the OMP algorithm is given by
\begin{align}
\mathcal{C}_{OMP} &\approx 2PMTW + \sum_{j=1}^{P} (\frac{j}{3}+ MT)j^2 + \sum_{j=1}^{P} 2jMT \\
&= \left(2PW + \frac{P^4}{12} + \frac{5}{18}P^3 + \frac{47}{36} P^2 + P \right)MT.
\end{align}
Since these operations are performed $KL$ times for the channel estimation, the complexity of the CS-based channel estimation $\mathcal{C}_{CS}$ is
\begin{align}
\mathcal{C}_{CS} = \left(2PW + \frac{P^4}{12} + \frac{5}{18}P^3 + \frac{47}{36} P^2 + P \right)KLMT.
\end{align}

We next analyze the complexity of CNN-based channel estimation~\cite{dong2019deep}.
Noting that the first convolutional output is obtained by multiplying the $C_1$ different $C_2 \times C_2$ filters to the real-valued LS channel estimate, the complexity is 
\begin{align}
\mathcal{C}_{conv_{1}} = 2 \cdot (C_{1} N_{R} N_{T}) \cdot 2C_{2}^{2} = 4C_{1} C_{2}^2 N_{R} N_{T}. \label{eq:CNNcomplexity1}
\end{align}
Let $C_3$ be the number of hidden layers in the CNN, then the complexity of the rest layers is
\begin{align}
    \mathcal{C}_{conv_{2}} = 2C_{1}^{2} C_{2}^{2} C_{3} N_{R} N_{T} + 4C_{1} C_{2}^{2} N_{R} N_{T}.
    \label{eq:CNNcomplexity3}
\end{align}
Noting that the aforementioned operations are performed $KL$ times for the channel estimation, we have the following complexity of the CNN-based channel estimation:
\begin{align}
\mathcal{C}_{CNN} = (8C_{1} C_{2}^{2} + 2C_{1}^{2} C_{2}^{2} C_{3}) KL N_{R} N_{T}.
\end{align}

\end{appendices}

\ifCLASSOPTIONcaptionsoff
  \newpage
\fi

\bibliographystyle{IEEEtran}

\begin{IEEEbiography}[{\includegraphics[width=1in,height=1.25in,clip,keepaspectratio]{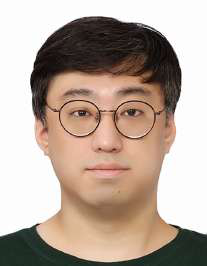}}]{Jinhong Kim} (jinhong3.kim@samsung.com)
received the B.S. degree from the Department of Electrical and Information Engineering, Seoul National University of Science and Technology, Seoul, Korea, in 2016 and the Ph.D. degree in Electrical and Computer Engineering, SNU, Seoul, Korea, in 2023. 
Since September 2023, he has been with Samsung Electronics, where he is working for 5G and 6G modem design. 
His research interests include sparse signal recovery and deep learning techniques for the 6G wireless communications.
\end{IEEEbiography}

\begin{IEEEbiography}[{\includegraphics[width=1in,height=1.25in,clip,keepaspectratio]{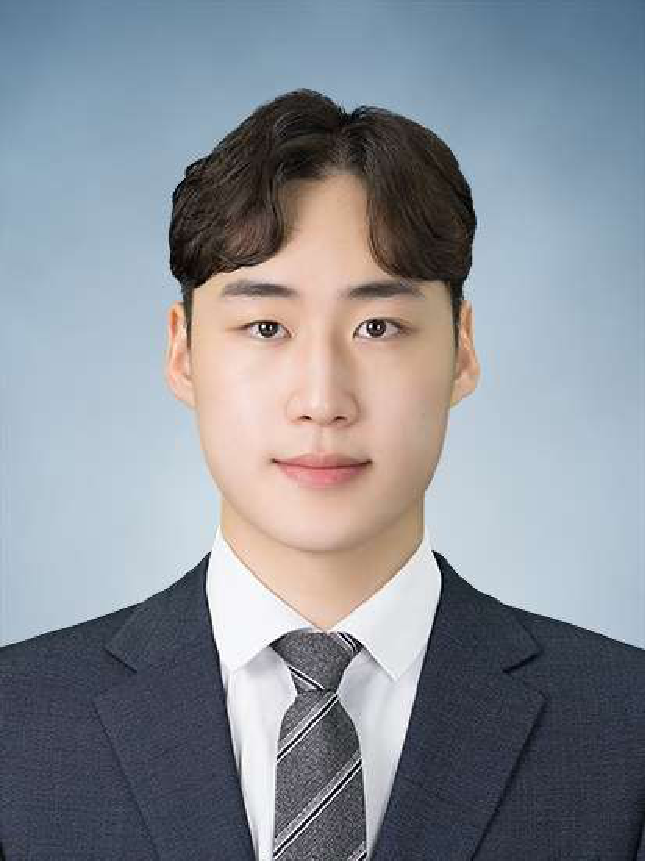}}]{Yongjun Ahn} (yong\_jun.ahn@samsung.com)
received the B.S. degree from the Department of Electrical and Computer Engineering, Seoul National University (SNU), Seoul, Korea, in 2019, where he received the Ph.D. degree in 2024.
Since March 2024, he has been with Samsung Electronics, where he is working for the development of 6G communication technologies. 
His research interests include 6G wireless communications and deep learning techniques.
\end{IEEEbiography}

\begin{IEEEbiography}[{\includegraphics[width=1in,height=1.25in,clip,keepaspectratio]{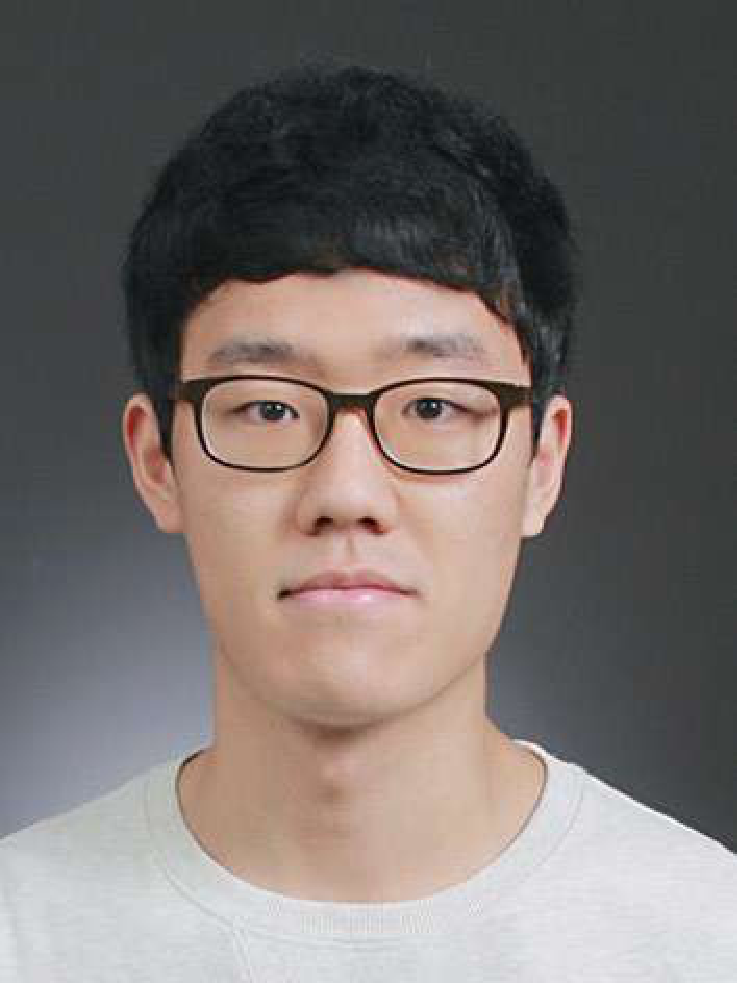}}]{Seungnyun Kim} (snkim94@mit.edu)
received the B.S. degree from the Department of Electrical and Computer Engineering, Seoul National University (SNU), Seoul, Korea, in 2016, where he received the Ph.D. degree in 2023.
He is currently a postdoctoral researcher with the Wireless Information and Network Sciences Laboratory, Massachusetts Institute of Technology (MIT), Massachusetts, USA. 
His research interests include beyond 5G and 6G wireless communications, optimization, and machine learning.
\end{IEEEbiography}

\begin{IEEEbiography}[{\includegraphics[width=1in,height=1.25in,clip,keepaspectratio]{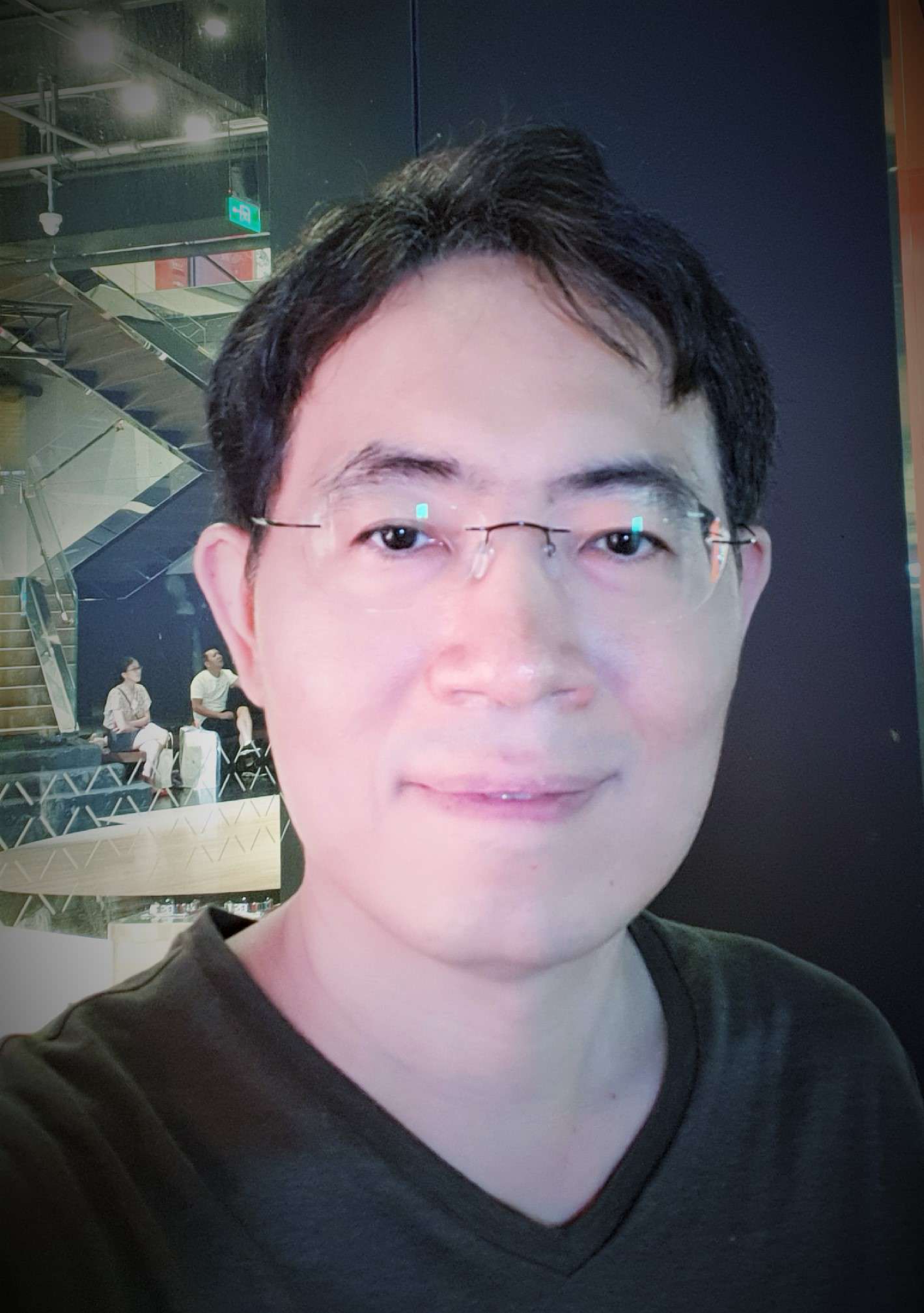}}]{Byonghyo Shim} (bshim@snu.ac.kr) received the B.S. and M.S. degrees in Control and Instrumentation Engineering from Seoul National University, South Korea, in 1995 and 1997, respectively, and the M.S. degree in mathematics and the Ph.D. degree in electrical and computer engineering from the University of Illinois at Urbana-Champaign (UIUC), Champaign, IL, USA, in 2004 and 2005, respectively.
From 1997 to 2000, he was an Officer (First Lieutenant) and an Academic full-time Instructor in the Department of Electronics Engineering, Korean Air Force Academy.
From 2005 to 2007, he was a Staff Engineer with Qualcomm Inc., San Diego, CA, USA.
From 2007 to 2014, he was an Associate Professor with the School of Information and Communication, Korea University, Seoul.
Since 2014, he has been with Seoul National University (SNU), where he is currently a Professor of the Department of Electrical and Computer Engineering and Vice Dean of the College of Engineering.
His research interests include wireless communications, statistical signal processing, and deep learning.
He was a recipient of the M. E. Van Valkenburg Research Award from the ECE Department, University of Illinois, in 2005, the Hadong Young Engineer Award from IEIE in 2010, the Irwin Jacobs Award from Qualcomm and KICS in 2016, the Shinyang Research Award from the Engineering College of SNU in 2017, the Okawa Foundation Research Award in 2020, the IEEE Comsoc AP Outstanding Paper Award in 2021, and the JCN Best Paper Award in 2024.
He was an Elected Member of the Signal Processing for Communications and Networking (SPCOM) Technical Committee of the IEEE Signal Processing Society.
He has been served as an Associate Editor for IEEE Transactions on Wireless Communications (TWC), IEEE Transactions on Communications (TCOM), IEEE Transactions on Vehicular Technology (TVT), IEEE Transactions on Signal Processing (TSP), IEEE Wireless Communications Letters (WCL), and Journal of Communications and Networks (JCN) and also a Guest Editor for IEEE Journal on Selected Areas in Communications (JSAC).
\end{IEEEbiography}

\end{document}